\documentclass[aps,prd,twocolumn,floats,balancelastpage,showpacs,preprintnumbers,floatfix,nofootinbib,superscriptaddress]{revtex4}

\usepackage{epsfig}
\usepackage{epstopdf}
\usepackage{float, nicefrac}
\usepackage[usenames]{color}
\usepackage{ulem}

\def\degr{\hbox{$^\circ$}}
\newcommand{\beq}{\begin{equation}}
\newcommand{\eeq}{\end{equation}}
\newcommand{\ben}{\begin{eqnarray}}
\newcommand{\een}{\end{eqnarray}}
\newcommand{\bi}{\begin{itemize}}
\newcommand{\ei}{\end{itemize}}
\newcommand{\hess}{H.E.S.S.~}
\newcommand{\g}{$\gamma$}

\begin{document}


\title{Constraints on axionlike particles with H.E.S.S. from the irregularity of the PKS~2155$-$304 energy spectrum}

\author{A.~Abramowski}
\affiliation{Universit\"at Hamburg, Institut f\"ur Experimentalphysik, Luruper Chaussee 149, D 22761 Hamburg, Germany}

\author{F.~Acero}
\affiliation{Laboratoire Univers et Particules de Montpellier, Universit\'e Montpellier 2, CNRS/IN2P3,  CC 72, Place Eug\`ene Bataillon, F-34095 Montpellier Cedex 5, France}

\author{F.~Aharonian}
\affiliation{Max-Planck-Institut f\"ur Kernphysik, P.O. Box 103980, D 69029 Heidelberg, Germany}
\affiliation{Dublin Institute for Advanced Studies, 31 Fitzwilliam Place, Dublin 2, Ireland}
\affiliation{National Academy of Sciences of the Republic of Armenia, Yerevan }

\author{F.~Ait Benkhali}
\affiliation{Max-Planck-Institut f\"ur Kernphysik, P.O. Box 103980, D 69029 Heidelberg, Germany}

\author{A.G.~Akhperjanian}
\affiliation{Yerevan Physics Institute, 2 Alikhanian Brothers St., 375036 Yerevan, Armenia}
\affiliation{National Academy of Sciences of the Republic of Armenia, Yerevan }

\author{E.~Ang\"uner}
\affiliation{Institut f\"ur Physik, Humboldt-Universit\"at zu Berlin, Newtonstr. 15, D 12489 Berlin, Germany}

\author{G.~Anton}
\affiliation{Universit\"at Erlangen-N\"urnberg, Physikalisches Institut, Erwin-Rommel-Str. 1, D 91058 Erlangen, Germany}

\author{S.~Balenderan}
\affiliation{University of Durham, Department of Physics, South Road, Durham DH1 3LE, U.K.}

\author{A.~Balzer}
\affiliation{DESY, D-15735 Zeuthen, Germany}
\affiliation{Institut f\"ur Physik und Astronomie, Universit\"at Potsdam,  Karl-Liebknecht-Strasse 24/25, D 14476 Potsdam, Germany}

\author{A.~Barnacka}
\affiliation{Nicolaus Copernicus Astronomical Center, ul. Bartycka 18, 00-716 Warsaw, Poland}

% ...

\collaboration{The H.E.S.S. Collaboration}
\noaffiliation

\begin{abstract}
Axionlike particles (ALPs) are hypothetical light (sub-eV) bosons predicted in some extensions of the Standard Model of particle physics. 
In astrophysical environments comprising high-energy gamma rays and  turbulent magnetic fields, the existence of ALPs can modify the energy spectrum of the gamma rays for a sufficiently large coupling between ALPs and photons. 
This modification would take the form of an irregular behavior of the energy spectrum in a limited energy range. Data from the H.E.S.S. observations of the distant BL Lac object PKS~2155$-$304  ($z = 0.116$) are used to derive upper limits at the 95\% C.L. on the strength of the ALP coupling to photons, $g_{\gamma a} < 2.1\times 10^{-11}$  GeV$^{-1}$ for an ALP mass between 15 neV and 60 neV. The results depend on assumptions on the magnetic field around the source, which are chosen conservatively. The derived constraints apply to both light pseudoscalar and scalar bosons that couple to the electromagnetic field.
\end{abstract}


\maketitle


\section{Introduction}

Some extensions of the Standard Model of particle physics predict the existence of pseudoscalar particles with sub-eV mass. A well-known example is the axion, originally introduced as a potential explanation of the absence of $CP$ violation in quantum chromodynamics (this is the so-called ``strong $CP$ problem"). The predicted particle is the axion, which is a pseudo Nambu-Goldstone boson associated to the spontaneous breaking of a U(1) symmetry (the Peccei-Quinn symmetry) at an energy scale $f$~\cite{Peccei:1977hh, Wilczek:1977pj, Weinberg:1977ma}. The original Peccei-Quinn model placed $f$ at the level of the electroweak (EW) scale and induced an axion mass of the order of 100 keV. This has been ruled out soon after studying the decays of quarkonia and the effect of axions on stellar evolution (see for instance~\cite{Kim:1986ax,1990PhR...198....1R}). Later it has been assumed that the scale $f$ was much larger than the EW scale, leading to a very light and weakly interacting axion called the ``invisible axion." Axions are predicted to couple to photons through a term containing $g_{\gamma a}\times a$, where $g_{\gamma a}$ is the photon-axion coupling constant (expressed in $\rm GeV^{-1}$) and $a$ the axion field. For the conventional axions, the coupling to photons $g_{\gamma a}$ and the axion mass $m$ are related as they are both proportional to $1/f$. The mechanism that leads to axions is, however, very generic and many models actually predict the spontaneous breaking of a global U(1) symmetry at high energy, resulting in the prediction of axionlike particles (ALPs, see for instance~\cite{Kim:1986ax}). ALPs can couple to photons in the same way as axions, but unlike axions their coupling strength  and mass are  generally independent parameters. For example, ALPs are ubiquitous in string theory, for which $f$ can be of order of the string scale and $m$ can be as low as $10^{-13}\;\rm eV$~\cite{Svrcek:2006yi, Arvanitaki:2009fg}. In some regions of the parameter space, even at these very low masses, ALPs are also good candidates for cold dark matter of the Universe~\cite{Arias:2012az}. They could have been produced by different mechanisms in the early Universe, either thermally or nonthermally~\cite{Arias:2012az}.

The interaction term between photons and ALPs can be written in terms of the electric field $\vec{E}$ and the magnetic field $\vec{B}$ as 
\begin{equation}
\mathcal{L}_{\gamma a}\;\; = \;\;g_{\gamma a}\;\vec{E}\cdot\vec{B}\;a\;\;.\label{eq:lagrangian2}
\end{equation}
This coupling opens up the possibility of oscillations between photon and ALP states in an external magnetic field~\cite{1988PhRvD..37.1237R} and enables experimental searches for ALPs. Four types of experiments are sensitive to ALPs (see~\cite{1983PhRvL..51.1415S} or~\cite{Beringer:1900zz} for a comprehensive review). The photon-ALP coupling is used to search for ALPs supposedly thermally produced in the Sun, as done with the CAST experiment~\cite{2011PhRvL.107z1302A}. In CAST, a magnet is pointed towards the Sun, with the intent to detect x rays from the conversion of ALPs into photons inside the apparatus. Another search strategy assumes that ALPs make up the cold dark matter and use resonant microwave cavities, like in the ADMX experiment~\cite{Asztalos:2009yp}. High intensity laser beams in magnetic fields are used to perform light-shining-through-a-wall type of experiments, as done for example in the ALPS experiment~\cite{Ehret:2010mh}. As a general rule, the efficiency of the photon-ALP oscillation mechanism in an external magnetic field is maximized for large values of the magnetic field and long propagation baselines, as both these parameters increase the probability of conversion from one state to another. Astrophysical environments can offer bright sources of photons, a wide range of magnetic fields and very long baselines. It is then natural to try to use astrophysics to search for ALPs. Each of these search strategies probe different regions of the parameter space, as summarized in~\cite{Beringer:1900zz} (see~\cite{2010ARNPS..60..405J} for an extensive review).

The very high energy gamma-ray sky is a promising place to search for ALPs \cite{2007PhRvL..99w1102H,2007PhRvD..76b3001M,2007PhRvD..76l3011H,2008PhLB..659..847D}. A widely discussed observable is the opacity of the Universe to gamma rays, due to pair production on photons of the extragalactic background light (EBL; see~\cite{1967PhRv..155.1408G,2013APh....43..112D}) which would be modified by the presence of ALPs \cite{2003JCAP...05..005C, 2007PhRvD..76l1301D, 2008PhRvD..77f3001S, Mirizzi:2009aj, 2009PhRvD..79l3511S, 2009MNRAS.394L..21D, 2011PhRvD..84j5030D,2013PhRvD..87c5027M}. In the present article an alternative approach is considered. A common feature of astrophysical magnetic fields is turbulence. It is shown in~\cite{Wouters:2012qd} that if photon-ALP oscillations occur in a turbulent magnetic field, the random nature of the field translates into irregularities in the observed energy spectrum of the source. For a given source, the level of irregularity depends on the coupling $g_{\gamma a}$ and the ALP mass $m$. The same kind of effect is pointed out in~\cite{Mirizzi:2005ng} in the case of quasar light absorption but never led to a constraint because of the highly irregular nature of the observed quasar spectra due to the Lyman-$\alpha$ forest.  Here, it is proposed to use a well observed gamma-ray source and measure the level of irregularity of its energy spectrum, to estimate the level of ALP-induced irregularity the data can accommodate, thus constraining the ALP parameter space.

In the next section, a short review of the relevant ALP phenomenology is given. The modeling of the magnetic fields on the line of sight from the gamma-ray source PKS~2155$-$304 is discussed in Sec.~\ref{sec:mag}.
In Sec.~\ref{sec:data}, the H.E.S.S. experiment is presented together with the data set. Section~\ref{sec:method} describes the estimation of the level of irregularity in the data, which is afterwards used to derive the limits in the $(g_{\gamma a},\,m)$ plane. The final constraints are shown and discussed in Sec.~\ref{sec:constraints}.

\section{Phenomenology of the photon-ALP system}
\label{sec:signal}

Because of the coupling given in Eq.~(\ref{eq:lagrangian2}), the photon-ALP system propagates as a mixing of three quantum states. Two states correspond to the photon polarizations and one state corresponds to the ALP. The propagation of the photon-ALP system is described as in \cite{Mirizzi:2009aj} with the formalism of the density matrix. The source beam is considered as unpolarized. This is correctly accounted for with an initially diagonal density matrix with equal probabilities of 1/2 for the two polarization states and null probability for the ALP state. The probability of observing a given state after traversing a region of size $s$ containing a coherent magnetic field of strength $B$ oscillates with a spatial wavelength 
\begin{equation}
\lambda_{\rm osc}\;\;=\;\;\frac{4\pi}{\sqrt{\Delta_{\rm a}^2 + 4\Delta_{\rm B}^2}}\;\;, 
\label{eq:lambda}
\end{equation}
with $\Delta_{\rm a} = -m^2/\left (2E \right )$ and 
$\Delta_{\rm B} = g_{\gamma a} B\,\sin \left ( \theta \right )/2$. 
Here $E$ is the energy of the photon-ALP system and $\theta$ accounts for the angle between the direction of the magnetic field and the axis of propagation. The magnetic fields are expressed in Lorentz-Heaviside units (1 G = $1.95\times10^{-20}$ GeV$^{2}$). The contribution from the electron plasma, that would modify the ALP mass term, is small compared to this latter and is hence neglected (relative contribution to $\Delta_{\rm a}$ less than $10^{-6}$). The birefringence induced by QED vacuum effects can also be neglected because of the too small magnetic fields involved (relative contribution less than $10^{-3}$). Finally, at TeV energies, Faraday rotation of the polarization axis in an external magnetic field is also negligible~\cite{Mirizzi:2009aj}.
For a 1 TeV gamma ray, typical of H.E.S.S. observations, the oscillation length within a $\mu$G environment is about 40~kpc, assuming an ALP mass below a few $\rm \mu eV$ and a coupling $g_{\gamma a}\sim10^{-10}\;\rm GeV^{-1}$ at the limit of the CAST constraint~\cite{2011PhRvL.107z1302A}. In other words, if an astrophysical environment hosting a high-energy gamma-ray source contains $\mu$G level magnetic fields with kiloparsec coherence lengths, the gamma rays have a significant chance to convert into ALPs (and vice versa). The same is true for nG level intergalactic magnetic fields on spatial lengths of the order of a few~Mpc. 

The conversion can only occur efficiently for $4\Delta_{\rm B}^2 \gtrsim \Delta_{\rm a}^2$ yielding a critical energy
\begin{equation}
E_{\rm c}\;\; = \;\; \frac{m^2}{2g_{\gamma a} B \sin \theta}\;\;,
\label{eq:threshold}
\end{equation}
above which the mixing is strong. For $g_{\gamma a}\sim10^{-10}\;\rm GeV^{-1}$ , $m=20\;\rm neV$ and $B=1\;\rm \mu G$, one finds $E_{\rm c}\sim 100\;\rm GeV$, which is the order of magnitude of the energy threshold for H.E.S.S. To keep the same critical energy with a lower magnetic field of 1~nG, the ALP mass has to be lowered by a factor $10^{3/2}$ to about 1~neV. 

The oscillation length for the photon-ALP system is energy dependent around $E_{\rm c}$, for $\Delta_{\rm a} \sim 2\Delta_{\rm B}$. At higher energies, $\Delta_{\rm a}\ll\Delta_{\rm B}$, spatial oscillations occur but the oscillation length does not depend on the energy. This means that around the energy $E_{\rm c}$, damped oscillations appear in the measured energy spectrum (as shown in Fig.~1 of~\cite{Wouters:2012qd}). 

As mentioned earlier, astrophysical magnetic fields are usually turbulent and the gamma-ray beams from high-energy sources cross many coherent magnetic domains. In a simplified picture, turbulent magnetic fields can be considered as patches of coherent domains, in each of which the direction of the magnetic field is randomly oriented. This image, though simplified, helps visualizing the phenomenology of the conversion. From one domain to the next, the orientation of the magnetic field changes, and so do the amplitude and the period of the spectral oscillations. Note that in the analysis presented below, a more realistic model is adopted, in which a distribution of magnetic field modes, covering a wide range of spatial dimensions, is considered. When several uncorrelated domains are crossed, unrelated oscillation patterns mix up and result in an irregular and unpredictable absorption pattern for the gamma-ray beam, as discussed in Sec.~\ref{sec:mag} and illustrated in Fig.~\ref{fig:signal}. ALPs significantly mixing with gamma rays would therefore yield irregularities in a limited region of the energy spectrum, corresponding to about one decade around the critical energy. This was pointed out in~\cite{Wouters:2012qd} as a possible peculiar signature of ALPs in the gamma-ray energy spectra of some blazars. As noted in~\cite{Galanti:2013afa}, the magnitude of the effect depends on the assumptions regarding the initial polarization state of the photon beam. In the first analysis in~\cite{Wouters:2012qd} the ALP-induced irregularity was computed for an initially fully polarized beam. It was shown in~\cite{Galanti:2013afa} that the  irregularity signal persists in case of an unpolarized beam, although at a lower level of statistical  significance. For the present analysis, all predictions have been made for the specific source environment making conservative assumptions for unknown parameters. In particular, the initial photon beam has therefore been assumed to be unpolarized.

 \section{PKS~2155$-$304 as the beam provider}
 \label{sec:mag}
 
 \subsection{Choice of PKS~2155$-$304}
 As the signal would take the form of an irregular absorption pattern in the observed energy spectrum of a source, for a given energy resolution the most important requirement for source selection is a strong statistics base. Large statistics enable the building of an accurate energy spectrum and lead to a better sensitivity to irregularities. In addition, the source must be chosen such that the photon beam crosses turbulent magnetic fields. As discussed in the previous section, better constraints are expected to result from stronger magnetic fields. Moreover, the magnetic field must have a spatial extent much larger than its coherence length. This is required for the creation of large spectral irregularities from the superposition of spectral pseudo-oscillations caused by regions with coherent magnetic fields. 
 
One of the most powerful and well-observed extragalactic TeV gamma-ray emitters is PKS~2155$-$304~\cite{2005A&A...430..865A, 2005A&A...442..895A,2009ApJ...696L.150A,2007ApJ...664L..71A,2009A&A...502..749A,2010A&A...520A..83H}. This BL Lac type active galactic nucleus is located at redshift $z=0.116$, offering the possibility for conversions in the intergalactic magnetic field (IGMF).  Blazars often reside in poor galaxy clusters of Mpc scale~\cite{1995ApJ...441..113S}. This is also the case for PKS~2155$-$304 which is at the center of a galaxy cluster of angular size of $5.7\pm0.5^\prime$~\cite{Falomo:1993dv}. Using a spatially flat $\Lambda$CDM universe with $\Omega_\Lambda=0.685$, $\Omega_{\rm m}=0.315$ and $H_0=67.3\;\rm km/s/Mpc$~\cite{2013arXiv1303.5076P}, the radius of the galaxy cluster lies in a range from 340 to 400 kpc. A value of 370 kpc is used in the following. Galaxy clusters of this class contain turbulent magnetic fields that are well characterized~\cite{2002ARA&A..40..319C}. Note that PKS~2155$-$304 is suggested as a good target for ALP searches based on opacity studies at high energies~\cite{Horns:2012kw}.

\subsection{General considerations on astrophysical magnetic fields}
Magnetic fields in galaxy clusters can be probed by measuring the rotation of the polarization of a radio photon beam due  to the Faraday effect (see \cite{Ryu:2011hu} for a recent review). Faraday rotation measurements in galaxy clusters show evidence for magnetic field strengths between 1 and 10~$\mu$G  with coherent modes on length scales ranging from 0.1 to 10~kpc (see ~\cite{2002ARA&A..40..319C, Ryu:2011hu} for reviews). The knowledge on the magnetic field in filaments and voids, {\it i.e.} IGMF, is much scarcer. Faraday rotation measurements are hard to perform since the background synchrotron emission is faint and uncertain, and because of the contamination from the Galactic foreground contribution \cite{Ryu:2011hu}. A coherence length of the order of 1~Mpc can be assumed for the IGMF~\cite{Akahori:2010ym}. Lower limits on its strength ranging from $10^{-18}$~G to $10^{-15}$~G are derived from searches for GeV gamma-ray emission from electromagnetic cascades from TeV blazars~\cite{2010Sci...328...73N,2011ApJ...733L..21D,2011A&A...529A.144T}. Note that authors of~\cite{Arlen:2012iy} argue that current observations are compatible with a zero IGMF hypothesis. The theoretical basis of such approaches also remains under debate~\cite{2012ApJ...752...22B,2012ApJ...758..102S}. Current upper limits on Mpc scales are close to 1~nG~\cite{1999ApJ...514L..79B,2013A&ARv..21...62D}. The turbulence of the magnetic field in the galaxy cluster is described by accounting for the power distribution of the modes. It is modeled in this work as in \cite{1999ApJ...520..204G} by a Gaussian random field with zero mean and a squared rms intensity $\delta B^2$ following a power spectrum:
\begin{equation}
\delta B_k ^2\;\propto\; \sigma^2\frac{k^2}{1+(kL_c)^{\alpha+2}}\;\;,
\label{eq:turb}
\end{equation}
 where $k$ is the wave number, $\sigma$ is the rms intensity and $L_c$ is the coherence length. This spectrum corresponds to a $k^{-\alpha}$ power law at large $k$.
 
 Such a description is more accurate than the mere assumption of a turbulence with only one scale, in which case the path is divided into cells of size equal to the coherence length, and where the orientation of the magnetic field is randomized in cell transitions.
The magnetic field turbulence can also be probed by Faraday rotation measures. This is done for instance in \cite{2010A&A...513A..30B,2005A&A...434...67V} for the Coma and Hydra galaxy clusters where a power spectrum compatible with a Kolmogorov slope $\alpha = 5/3$ is found on scales from 0.1 to 10~kpc. 

\subsection{Magnetic field along the PKS~2155$-$304 line of sight}

As the magnetic field in the galaxy cluster around PKS~2155$-$304 is not measured, it is necessary to make some assumptions.
It is frequent that galaxy clusters are observed around radio galaxies. As stated before, the magnetic field strength and structure is well characterized in some prototypic objects. Blazars like PKS~2155$-$304 belong to the same family of objects~\cite{1993AJ....106..875L}, they are radio galaxies for which the jet points towards the Earth. As a galaxy cluster is actually observed around PKS~2155$-$304, it is reasonable to assume a magnetic field similar to that observed for other radio galaxies in clusters. Concerning the strength of the magnetic field in the cluster, the most conservative assumptions are made. In addition, the constraints on the ALP parameters are presented in~Sec.~\ref{sec:constraints} in a way independent of the magnetic field strength before they are converted using the conservative value. Concerning the structure of the magnetic field, in the following, a magnetic field turbulence index $\alpha = 5/3$ is assumed for the magnetic field turbulence index in the galaxy cluster of PKS~2155$-$304, with a maximal turbulence scale $L_c=10$~kpc. In Sec.~\ref{sec:constraints}, it is shown how the sensitivity of the analysis varies when these parameters are changed within a reasonable range. As shown in \cite{Wouters:2012qd}  that small turbulence scales rapidly become irrelevant for photon-ALP conversion. Hereafter, conversions within the galaxy cluster of PKS 2155$-$304 on scales lower than $L_c/10 = 1$ kpc are neglected with respect to those on the largest scales. Magnetic field strengths between 1 and 10~$\mu$G can be reasonably assumed in the case of the galaxy cluster embedding PKS~2155$-$304~\cite{2002ARA&A..40..319C}. A conservative value of the rms intensity of the field of 1~$\mu$G is considered in the following for deriving the limits.  Concerning the IGMF description, it is not obvious whether a description including turbulence is correct. Small scale perturbations could be damped by dissipative processes such as photon or neutrino diffusion so that the use of a Kolmogorov spectrum would be irrelevant~\cite{2001PhR...348..163G}. Throughout this article, the turbulence is modeled on a single scale of 1 Mpc. This description corresponds to the simple cell model with Gaussian distribution of the magnetic field strength. In the following, an rms intensity of 1~nG is used for the IGMF strength implying that the constraints are derived according to the most optimistic model for the IGMF, leading to less conservative constraints. The cluster magnetic field and the IGMF being of widely separated strength, they produce irregularities in different energy ranges for a given ALP mass. Therefore, constraints from the mixing in the galaxy cluster magnetic field and in the IGMF can be derived independently. 

Before detection on the Earth, the entangled photon-ALP system traverses the Galactic magnetic field which has a turbulent component with an rms intensity of a few $\rm \mu G$ on scales smaller than 1 kpc. This magnetic field strength is similar to that of galaxy clusters, but on smaller scales. As shown in~\cite{2007PhRvD..76b3001M}, because of the large number of domains, the conversion does not happen on such small turbulence scales. For that reason the Galactic magnetic field can be ignored for the present case. Even if the Galactic magnetic field was such that it would induce a small additional irregularity signal, not considering it here is a conservative approach.

\begin{figure}
\centering
\includegraphics[width=\columnwidth]{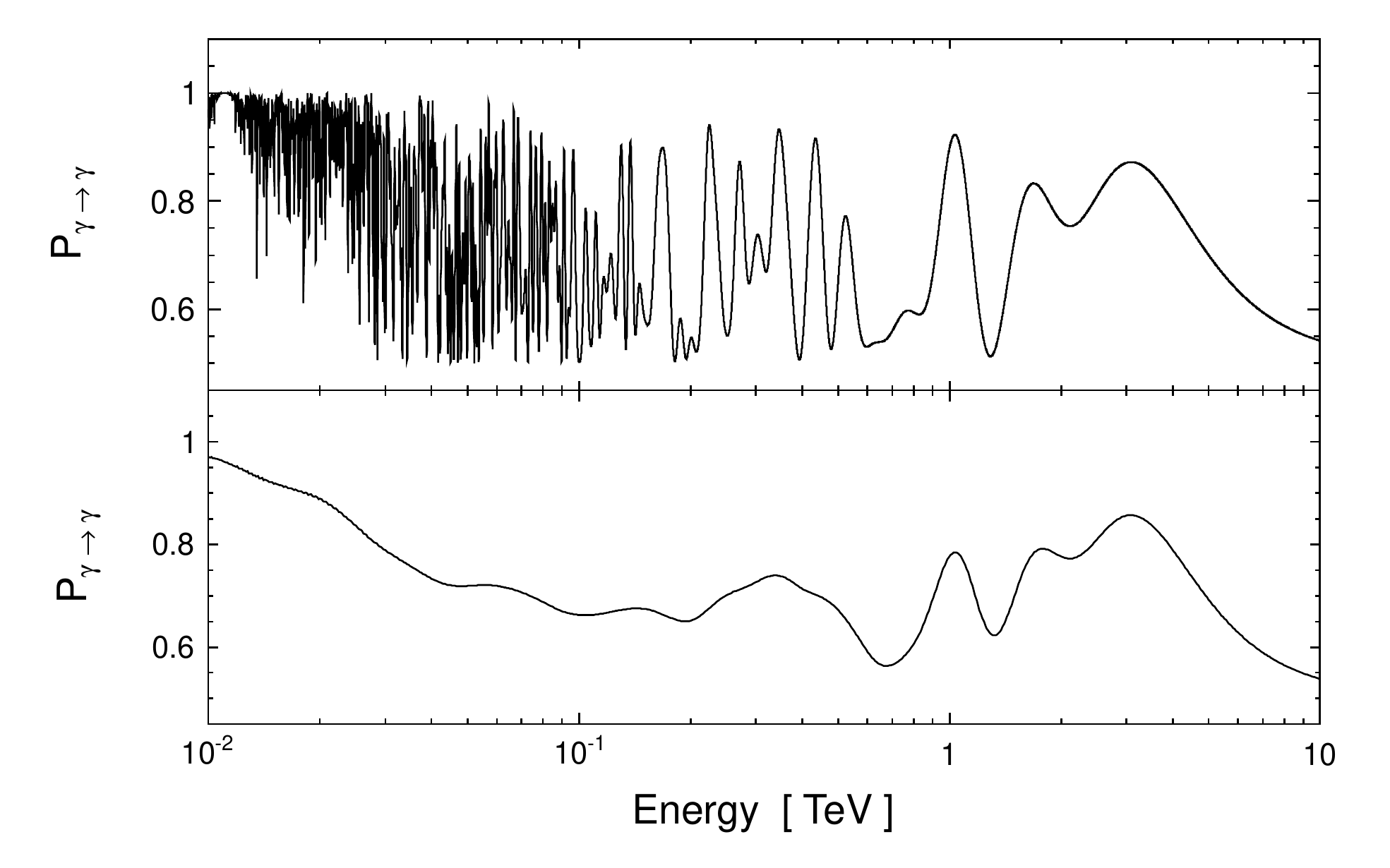}
\caption{Survival probability for gamma rays mixing with ALPs in a galaxy cluster magnetic field (see text for details). Top panel: Raw function. Bottom panel: The same function convolved with the energy resolution and bias of H.E.S.S. The instrumental response functions at 50 GeV are extrapolated to lower energies, not reachable with H.E.S.S.
 \label{fig:signal}}
\end{figure}

To help visualizing the type of signal being searched for, Fig.~\ref{fig:signal} shows an example of a typical modulation function for PKS~2155$-$304 in an energy range from 10~GeV to 10~TeV. This modulation function is called the survival probability and is the probability that for an incoming photon, a photon is measured in the end. Here, because the initial beam is assumed to be unpolarized, the photon survival probability cannot be lower than 0.5. For this prediction, $m=30\;\rm neV$, $g_{\gamma a}=10^{-10}\;\rm GeV^{-1}$ are used as ALP parameters. A cluster magnetic field of 1~$\mu$G is considered over a distance of 370~kpc with a coherence length of 10~kpc. The upper panel of Fig.~\ref{fig:signal} displays the raw expectation and the lower panel displays the same prediction convolved with the H.E.S.S. energy resolution and bias in the case of the selected observation of PKS~2155$-$304 (the instrument response functions are discussed in Sec.~\ref{sec:data}). This signal results from one possible realization of the turbulent magnetic field. Whereas the exact shape of the spectrum cannot be predicted, the statistical properties of the signal --and in particular the variance of the irregularity-induced noise-- are a prediction of the model, depending only on the mixing angle~\cite{Wouters:2012qd}. For this reason, Monte Carlo simulations are performed to obtain the constraints on a statistical basis. Sets of parameters that have a high probability to produce irregularity at a larger level than observed will be excluded.

\section{\hess data set}
\label{sec:data}

\begin{figure}[t]
\centering
\includegraphics[width=\columnwidth]{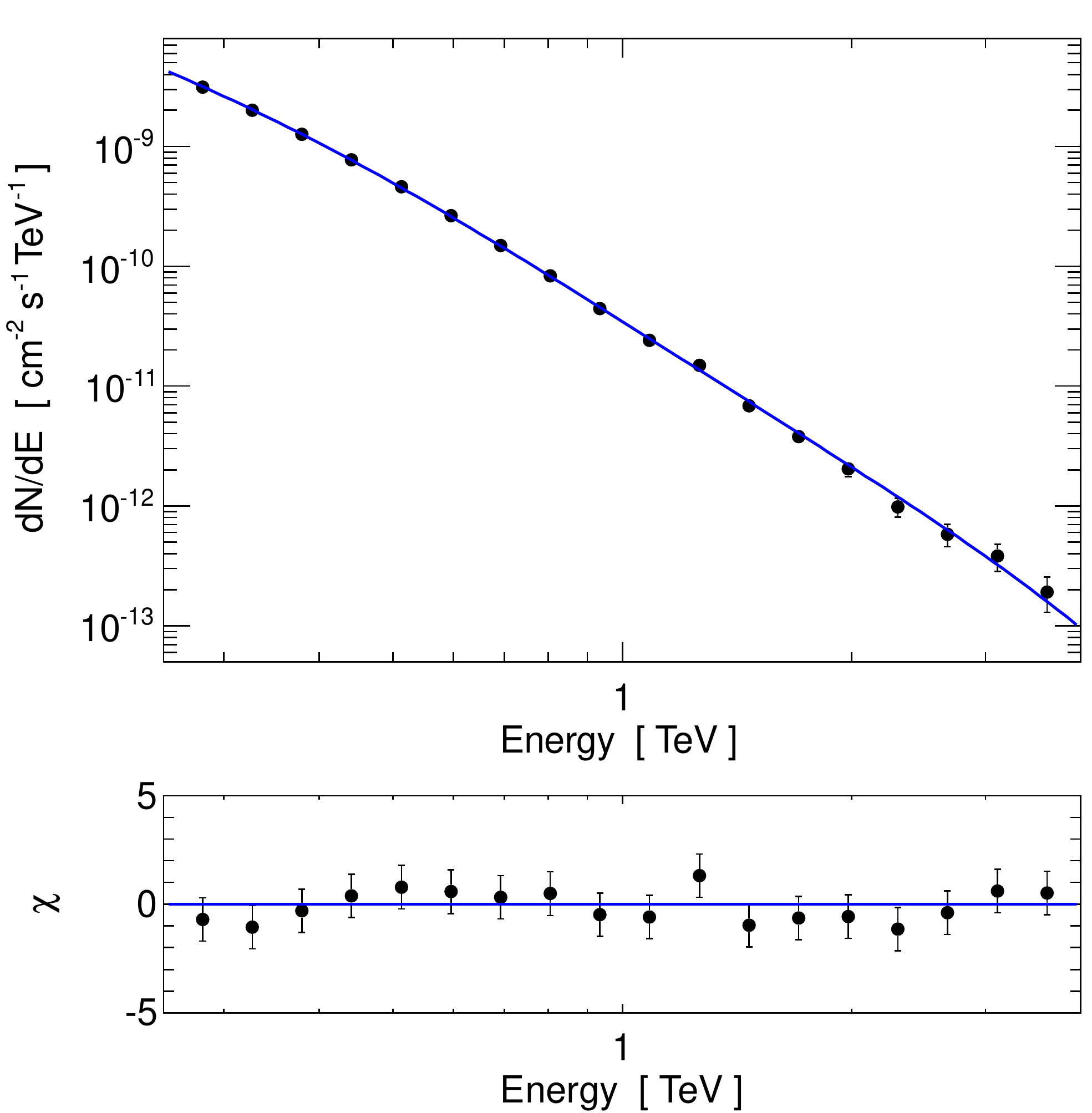}
\caption{Time-averaged energy spectrum of PKS~2155$-$304 for the data set used in the analysis. Top panel: The blue line is the best fit of a log-parabola modulated by absorption on the EBL to the data. Bottom panel: Relative residuals of the fit normalized to the errors.}
\label{fig:spectrum}
\end{figure}

PKS~2155$-$304 has been observed with H.E.S.S. and a large data set is available on that source, making it a good candidate to be used for deriving constraints on ALPs. \hess is an array of five imaging atmospheric-Cherenkov telescopes that are used to observe TeV \g rays and are situated in the Khomas highland of Namibia. During the first phase of the project from 2003 to 2012, four telescopes of 12 m diameter observed the \g-ray sky above a typical energy threshold of a few 100 GeV, the exact value depending on the observation conditions. A fifth 28 m diameter telescope started operation in 2012 with the aim of lowering the energy threshold down to tens of GeV. The used data set has been taken during the four telescope phase. More details about the first phase of \hess can be found in \cite{2006A&A...457..899A}.

The data set used in this paper is chosen to optimize the signal over noise ratio, to obtain the most accurate spectrum possible, in particular at high energy. It is based on observations taken in 2006, between the 27th of July and the 1st of August, when the source was in a high state \cite{2007ApJ...664L..71A}. 
The observed flux for this period is highly variable, ranging in a factor from 1 to more than 20. This is not a concern for this analysis since the irregularity effect is independent of the spectrum, so the averaged spectrum can be solely considered. 
Observations were taken in a large range of zenith angles from 5\degr\ to 45\degr\, ensuring both a low energy threshold of 250~GeV and a high effective area at energies above 1~TeV. A pointing offset from PKS 2155$-$304 of 0.5$^{\circ}$ is maintained in order to simultaneously evaluate the signal and the background from the same field of view. After data quality selection and dead-time correction, a total of 13~h of high quality data are used in the spectral analysis. The data are analyzed with the \textit{Model} analysis~\cite{deNaurois:2009ud},  in which a semianalytical model of electromagnetic air showers is used to fit the images recorded by the cameras. Loose selection criteria are applied for selection of the events, resulting in a low energy threshold for the spectrum reconstruction. The analysis is cross-checked with an independent calibration and analysis chain \cite{2006A&A...457..899A} giving consistent results. The instrument response functions are obtained from Monte Carlo simulations. The average energy resolution is 12\%, and the bias in the energy reconstruction is lower than 2\% in the considered energy range from 250~GeV to 4~TeV.

The spectrum of the 45505 \g-ray candidates (46124 ON events, 6186 OFF events, background normalization 0.1) is reconstructed using an unfolding technique, as described in~\cite{Albert:2007qw}, with regularization by iterations. This regularization is chosen to minimize the interbin correlation. The covariance matrix determined during the unfolding procedure is used in all spectral analyses presented below in order to take into account the remaining correlations between bins. 
Figure~\ref{fig:spectrum} shows the averaged energy spectrum for the considered data set. The unfolding procedure allows one to quantify the level of irregularity in the spectrum without assuming a particular spectral shape. It has been checked that the results (both for the spectrum itself and the final constraints based on the measured irregularity level) are consistent with those obtained with the forward folding procedure \cite{2001A&A...374..895P} used in H.E.S.S. 
The spectrum found in this study is compatible with the spectrum measured during the nights of the big flares (MJD 53943, 53946)~\cite{2009A&A...502..749A,2010A&A...520A..83H}.

The spectrum is well described  ($\chi^2 /n_{\rm d.o.f.}$ = 8.0/15) by a log-parabola shape modulated by absorption on the EBL:
\beq
 \frac{\mathrm{d}N}{\mathrm{d}E} \propto \left(\frac{E}{1 \mathrm{TeV}}\right)^{-\alpha-\beta \log(E/1 \mathrm{TeV})}e^{-\tau_{\gamma\gamma}(E)}  \;\;,
 \eeq 
 with $\alpha = 3.18 \pm 0.03_{\rm stat} \pm 0.20_{\rm syst}$, $\beta = 0.32 \pm 0.02_{\rm stat} \pm 0.05_{\rm syst}$. The optical depth $\tau_{\gamma\gamma}$ describes the absorption of gamma rays on the EBL modeled as in~\cite{2008A&A...487..837F}.  The integrated flux above 200 GeV is $F(>200 \;\mathrm{GeV}) = (8.68 \pm 0.40_{\rm stat} \pm 1.30_{\rm syst})\times10^{-10} \rm cm^{-2} s^{-1}$.

A fit without EBL absorption does not give satisfactory results, with $\chi^2/n_{d.o.f.}=315/15$ for a fit with a log-parabola without the EBL effect. This corresponds to a positive detection of the EBL-induced absorption, in agreement with the H.E.S.S. results on the EBL density measurement~\cite{2013A&A...550A...4H}. Details about related systematics and dependence on the EBL model can be found in~\cite{2013A&A...550A...4H}. The distortion of the spectrum due to the EBL absorption is very different from the one sought from ALPs, as the EBL-induced wiggle is decade wide in energy, compared to bin-to-bin fluctuations in the case of the ALP signal. Note that in the following analysis, no assumption on the spectral shape is made. As a consequence, the results of the analysis do not depend on assumptions regarding the EBL model.

The residuals to the best fit are displayed in the lower panel of Fig.~\ref{fig:spectrum}. Each residual is divided  by the uncertainty on the corresponding point, thus showing directly the number of standard deviations from the fit. Should an irregular behavior be present in the energy spectrum, it would manifest itself through fluctuations in this residual distribution. At this point this conclusion is qualitative. In the next section, a method is developed that is sensitive to bin-to-bin fluctuations in the energy spectrum itself. That is a safer approach, as the residual distribution depends on an assumption regarding the spectral shape. The residuals are shown here for illustration purpose only

\section{Method}
\label{sec:method}

\begin{figure}[h]
\centering
\includegraphics[width=.7\columnwidth]{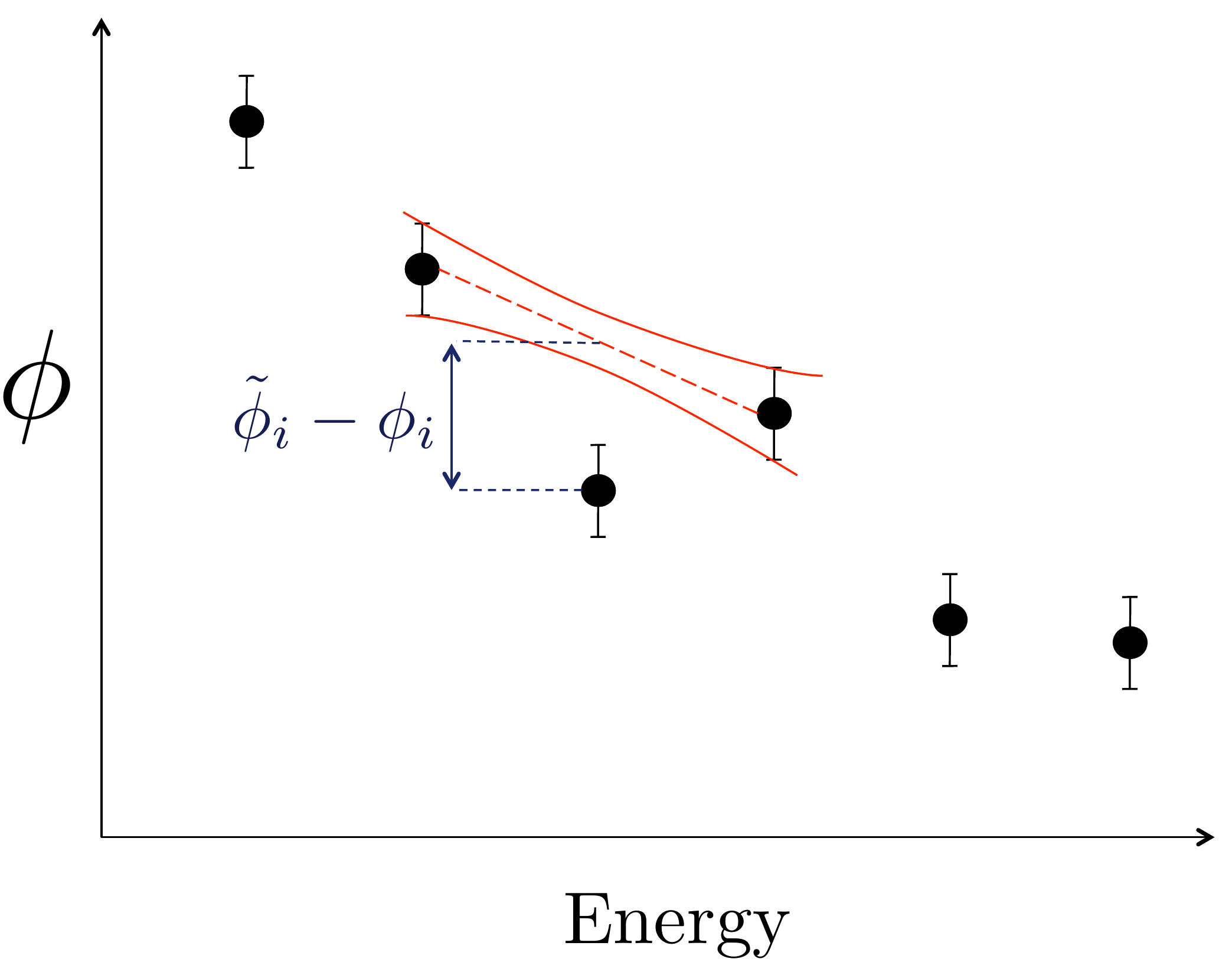}
\caption{Schematic view of the procedure used to quantify spectral fluctuations.\label{fig:estimator}}
\end{figure}

The method for deriving constraints aims at searching for the maximum level of irregularity induced by photon-ALP oscillations that  is allowed by the data once added to the regular shape of the spectrum. Although cosmic sources provide intrinsically smooth spectra at TeV energies, observed spectra naturally contain a certain amount of irregularity, due to the finite statistics, possible nontrivial -- yet unknown -- absorption features and instrumental responses. Given the shape of the typical ALP-induced signal, it is highly unlikely though that the irregularity would be compensated exactly by such effects. A discussion of the smoothness of the instrumental response follows in Sec.~\ref{sec:constraints}. In~\cite{Wouters:2012qd}, the variance of the residuals from a spectral fit is proposed as an irregularity estimator. However, for this estimator, an underlying smooth spectral shape has to be assumed, potentially introducing a bias. A more conservative approach makes use of an estimator that relies on minimal assumptions regarding the intrinsic spectrum. The estimator proposed here does not make use of a global fit but nevertheless assumes that the spectrum is locally well represented by a power law. The local power-law behavior is tested over the energy ranges of three consecutive bins of the spectrum displayed in Fig.~\ref{fig:spectrum}. On such narrow energy ranges, deviations from a power-law behavior are not expected in the framework of the underlying acceleration and radiation processes \cite{2001A&A...367..809K}. Each group of three consecutive bins is taken separately to form a triplet. In this way, $n-2$ triplets can be formed, where $n = 18$ is the number of bins in the spectrum.  Let $\phi_i$ being the measured flux in bin $i$ and $\tilde{\phi}_i$ the flux in the median bin expected from the power-law fit to the side bins. Then assuming a power-law interpolation on the two side bins, one has
\begin{equation}
\tilde{\phi}_i=\frac{\phi_{i+1}^{\beta_i}}{\phi_{i-1}^{\beta_i-1}}\;\;\text{with}\;\;\beta_i=\frac{\log\frac{E_{i-1}}{E_i}}{\log\frac{E_{i-1}}{E_{i+1}}}\;\;.
\end{equation}
For each triplet, the residual $(\tilde{\phi}_i-\phi_i)$ of the middle bin from the power law defined by the side bins is computed, as illustrated in Fig.~\ref{fig:estimator}. The residuals are normalized to account for the errors and correlations and then quadratically summed to form the irregularity estimator $\mathcal{I}$ defined by
\begin{equation}
\mathcal{I}^2= \sum_i {\frac{\left ( \tilde{\phi}_i - \phi_i \right )^2}{\vec{d}_i^{ \;T}C_i\vec{d}_i}}\;\;,
\label{eq:i}
\end{equation}
where $C_i = \mathrm{cov}(\phi_{i-1}, \phi_i, \phi_{i+1})$ is the covariance matrix for the triplet $i$ and 
\begin{equation}
\vec{d}_i\,^T =   \left(\frac{\partial \tilde{\phi}_i}{\partial\phi_{i-1}}, -1, \frac{\partial\tilde{\phi}_i}{\partial\phi_{i+1}}\right) \;\;.
\label{eq:vecd}
\end{equation}
In the absence of anomalous irregularities, the mean of $\mathcal{I}^2$ is the number of triplets that can be formed. This estimator is well suited to look for rapid fluctuations from bin to bin, which is a clear specificity of the expected signal. Effects that occur on wider energy ranges, like for instance the wiggle due to the EBL absorption~\cite{1999A&A...349...11A,2003A&A...403..523A,2013A&A...550A...4H}, should not contribute significantly to the measured irregularity level.

This estimator $\mathcal{I}$ can be used to constrain the ALP parameters $g_{\gamma a}$ and $m$ by estimating the expected level of irregularity. The random nature of the magnetic field implies that from one realization to another, the irregularity estimator does not take a single value. The expected signal distribution for different parameter sets is therefore determined from simulations of spectra. For each set of ALP parameters, $1000$ spectra are simulated with overall shape and statistics corresponding to the measured spectrum in Fig. 2. Each simulated spectrum is modified according to the expected photon-ALP oscillation for a randomly chosen magnetic field configuration (either inside the cluster or in the IGMF, depending on the choice of ALP parameters). The normalized distribution of the irregularity estimators for these simulated spectra is interpreted as a probability density function (PDF) for the ALP parameters under consideration. If the measured irregularity estimator is lower than 95\% of the simulated estimators, the corresponding ALP parameter set is considered as excluded at the 95\% C.L. 

\begin{figure}[t]
\centering
\includegraphics[width=0.9\columnwidth]{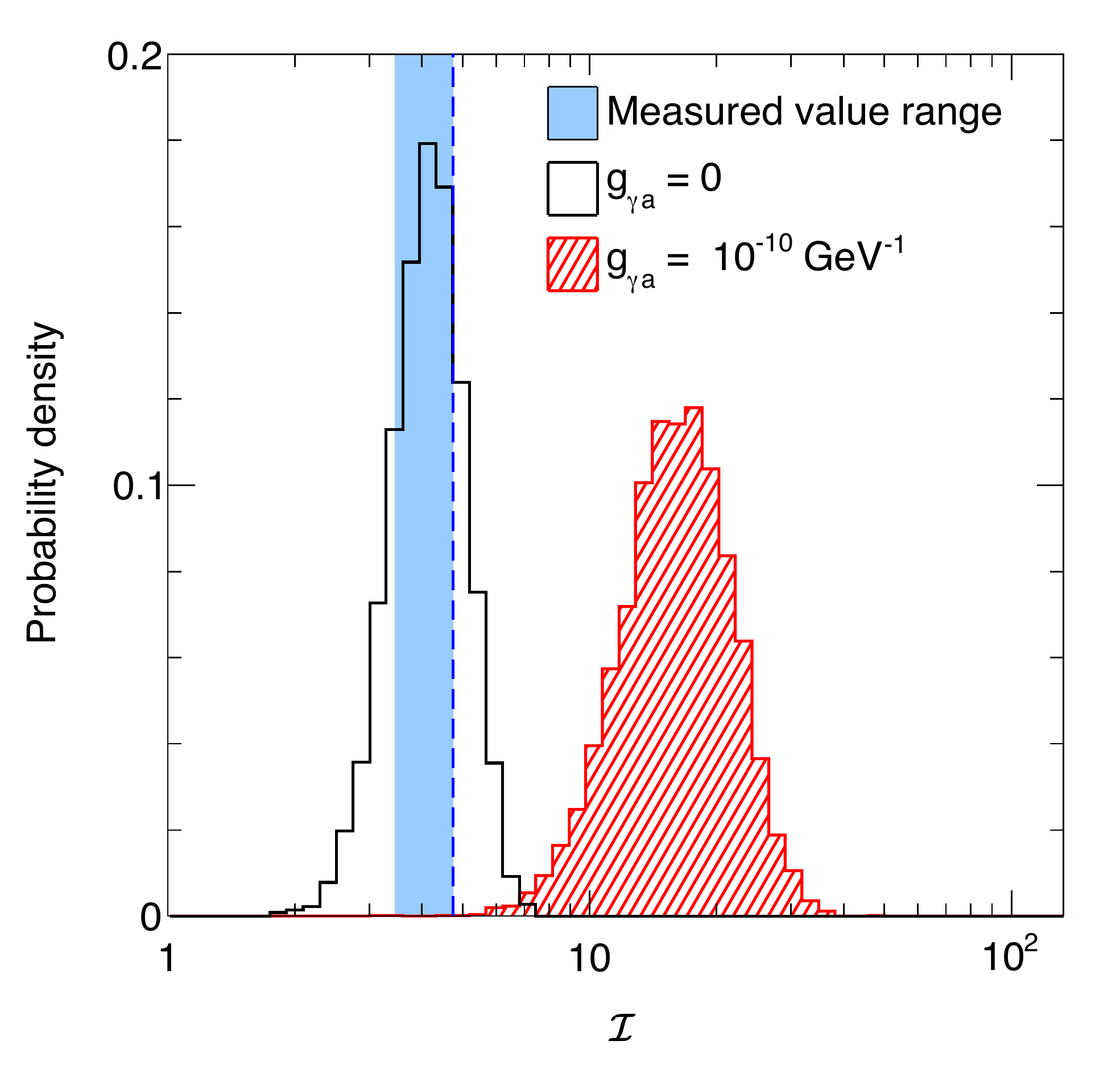}
\caption{Predicted probability density functions of the irregularity estimator for two ALP parameter sets. The vertical band corresponds to the rms of the fluctuations of the measurement when varying the binning. The dashed line indicates the value used to set the limits. \label{fig:dist}}
\end{figure}

One example of such a PDF is shown in Fig.~\ref{fig:dist} for $g_{\gamma a}= 10^{-10}$~GeV$^{-1}$. On the same figure, the distribution for a vanishing coupling $g_{\gamma a}=0$ is also shown, displaying the range of irregularity measurements one would obtain out of many realizations of the observation in the absence of ALPs. The vertical band corresponds to the measured irregularity in the data, as explained below, the width is due to the binning-related systematic error.

Two alternative irregularity estimators were tested, the variance of the residuals from a spectral fit to a smooth function, and the power spectrum density of the energy spectrum. The latter in principle measures the level of noise in the spectrum. Both gave results consistent with those derived from the estimator ${\mathcal I}$, although leading to slightly stronger exclusion limits, due to more stringent assumptions on the intrinsic spectral shape. The use of these two alternative estimators was abandoned in favor of the $\mathcal{I}$ estimator because of the conservativeness and the weaker dependence on spectral assumptions of the latter. Even when $\mathcal{I}$ is used, the limit has a weak dependence on the spectral shape assumed for the simulated spectra, that are used for building up the irregularity PDFs. It has been checked, however, that propagating the error on the spectral index has a negligible effect on the final exclusion limits. Furthermore, although the energy resolution from the full H.E.S.S. Monte Carlo is used in the simulations, an artificial modification of the energy resolution between 10\% and 20\% has only a marginal (less than 10\%) effect on the constraints.

As the irregularity is estimated from the variations of neighboring energy bins, a potential source of systematic error is the choice of the bin size. In order to check for possible systematics from the binning, the analysis is reproduced with different sizes for the bins. When the bin size is changed from $\Delta E/E=10\%$ to $\Delta E/E=20\%$, the measured level of irregularity is constant with a certain level of fluctuations, even when the total number of bins does not change. These fluctuations are due to the bin-to-bin reshuffling of the events during the rebinning procedure, either due to a change in the number of bins or to slight changes in the positions of the bin edges.
As a consequence one can consider the rms of the corresponding distribution as the systematic error on the value of $\mathcal{I}$ due to the choice of the bin size.  In Fig.~\ref{fig:dist}, the vertical band corresponds to the $1\sigma$ range for the measurement of $\mathcal{I}$ in the unfolded spectrum. The upper end of this interval (the vertical dashed line in Fig.~\ref{fig:dist}) is used in the following to set the limit.

\section{Constraints on the ALP parameters}
\label{sec:constraints}

The irregularity level measured from the unfolded spectrum shown in Fig.~\ref{fig:spectrum} is found to be $\mathcal{I}=4.10\pm 0.65$, where the error is the rms of the fluctuations of $\mathcal{I}$ when varying the binning. Following the prescription in Sec.~\ref{sec:method}, the value of $\mathcal{I}=4.75$ is used to derive limits from the irregularity PDFs, separately for the case of conversion in the cluster magnetic field and in the IGMF.

\begin{figure*}
\centering
\includegraphics[width=\textwidth]{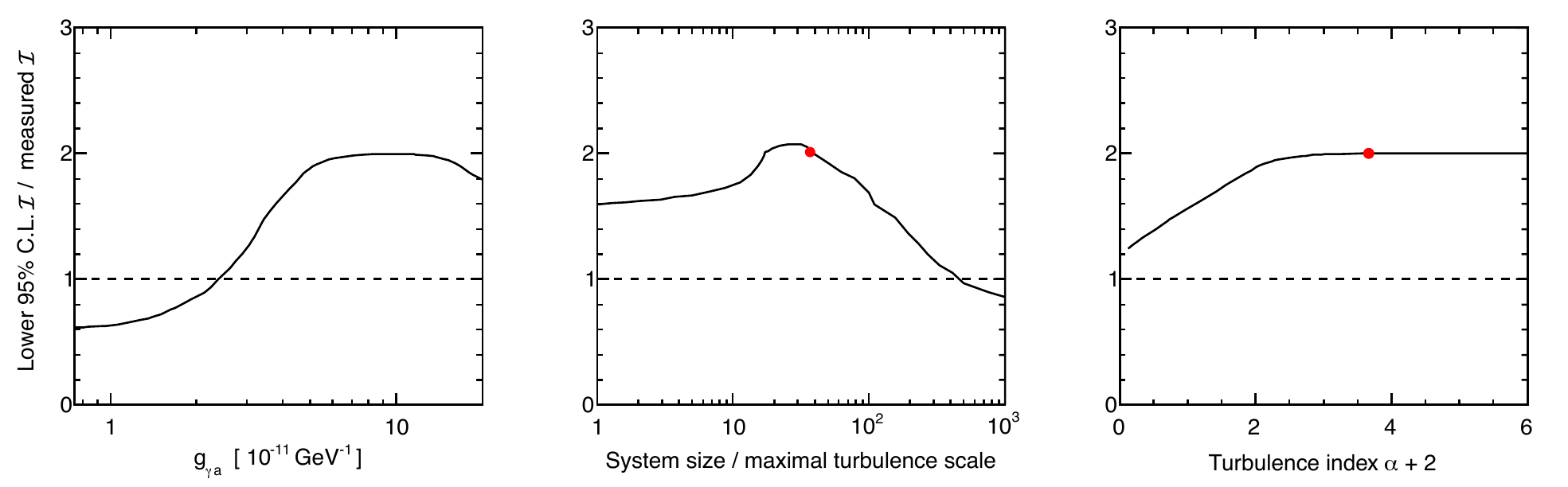}
\caption{Evolution of the one-tailed 95\% lower bound of the PDF constructed for the estimator normalized to its measured value as a function of the system parameters. Each panel shows variation of one parameter around the value $g_{\gamma a}\times B = 10^{-10} \,\mathrm{GeV}^{-1}\times1\mu \rm G$, $m = 30$~neV and conversion in the galaxy cluster with a turbulence power spectrum slope $\alpha=5/3$ on scales between 1 and 10 kpc. For the middle panel and the right panel, the value used in the analysis is represented by a marker. From left to right, variations of the coupling constant, of the maximal turbulence scale compared to the total system radius of 370 kpc (number of corresponding domains) and of the turbulence power spectrum slope are displayed.}
\label{fig:param}
\end{figure*}

The measured value of $\mathcal{I}$ indicates that the spectrum of PKS~2155$-$304 does not exhibit strong irregularities, meaning that limits can be established on the ALP parameters. The only loophole could be that ALP-induced irregularities would be compensated exactly by an unknown energy-dependent effect in the instrument or the analysis chain. The exclusion limits are derived on a statistical basis from the simulation of many irregularity pattern realizations. If, in one specific realization of the magnetic field configuration, the ALP signal was compensated by some unusual effect, it would not be the case in all other realizations. 
Assuming we live in this specific realization requires an extreme fine-tuning. This possibility can therefore be safely ignored.
To test the smoothness of the instrument response and ensure that irregularities in it cannot compensate ALP-induced ones, tests are performed on a control sample in which no irregularity signal is expected. The same procedure as for PKS~2155$-$304 is applied to the Crab Nebula data, from which no ALP signal is expected in the considered energy range. 
The Crab Nebula data set is chosen such that it offers statistics similar to the present analysis. Because of the larger zenith angle, the threshold is 600 GeV, and the spectrum is cut at 4 TeV to restrict it to the energy range covered by the measured spectrum of PKS~2155$-$304.
After correcting for the different number of bins in the energy spectrum, one finds $\mathcal{I}_{\rm Crab}=3.59\pm0.38$. That value is compatible with the one measured on the PKS~2155$-$304 energy spectrum, showing that this level of fluctuation is common for the H.E.S.S. observations.

For illustration, Fig.~\ref{fig:param} (left panel) displays the ratio of the 95\% C.L. lower limit 
for $\mathcal{I}$, normalized to the measured value $\mathcal{I}=4.75$, as function of the 
coupling strength $g_{\gamma a}$ for a fixed ALP mass of $m=30$~neV, assuming
conversion in the magnetic field of the galaxy cluster.
Since the 95\% C.L. lower bound of the PDF is normalized to the measured value, a value greater than 1 means
that the irregularity level is too high to be in agreement with the data at the 95\% C.L.
The ratio crosses unity at $g_{\gamma a}=2.1\times 10^{-11}\;\rm GeV^{-1}$, which hence
represents the 95\% C.L. upper limit on the photon-ALP coupling strength
for the considered ALP mass. Note that the ratio increases up to coupling strengths of 
$g_{\gamma a}\approx 10^{-10}\;\rm GeV^{-1}$ and then decreases again. This is due to the fact that irregularities appear only in the vicinity of
the critical energy $E_c$ given in Eq.~(\ref{eq:threshold}). For coupling strengths larger 
than $10^{-10}\;\rm GeV^{-1}$, the irregularities would move into energy ranges 
too low to be measurable by H.E.S.S. Since, according to Eq.~(\ref{eq:threshold}), $E_c$ depends on 
$m^2/g_{\gamma a}$, the sensitivity at larger coupling strengths is restored
for larger ALP masses. 

The PDFs of the estimator $\mathcal{I}$ are derived under well motivated assumptions for the
magnetic field configurations, as described in Sec.~\ref{sec:mag}. The
sensitivity of the irregularity measure $\mathcal{I}$ to these assumptions is studied
by varying the number of domains, corresponding to the ratio of system size $L$
and maximum turbulence scale $s$, and the slope $\alpha+2$ of the turbulence power 
spectrum in Eq.~(\ref{eq:turb}). Figure ~\ref{fig:param} (middle and right panels) displays the results for an ALP mass 
of $m=30$~neV and  a coupling strength of $g_{\gamma a}= 10^{-10}\;\rm GeV^{-1}$, 
at which the irregularity measure is close to its maximum for conversions
in the galaxy cluster. The irregularity
level is insensitive to the power spectrum slope over a wide parameter range.
In contrast, a strong dependence on the maximum turbulence scale is observed.
The sensitivity to irregularities is the largest for maximum turbulence scales 
corresponding to 20 domains of coherent magnetic fields (this value depends on the energy resolution of the instrument). 
At higher turbulence scales (smaller number of domains), turbulence is too scarce to
produce large irregularities, whereas at lower turbulence scales (larger
number of domains), spectral oscillation structures become too densely spaced
to be resolved within the finite energy resolution of the instrument. For the galaxy
cluster magnetic field, a maximum turbulence scale of 10 kpc (corresponding
to 37 domains) is well motivated, although other values between 5 and 10 kpc
are also reported~\cite{2002ARA&A..40..319C}. This happens to coincide with the scale where
the sensitivity to irregularities is at its maximum. This strengthens
the motivation to choose PKS~2155$-$304 for the search
for ALP-induced spectral irregularities.

\begin{figure*}
\centering
\includegraphics[width=0.8\textwidth]{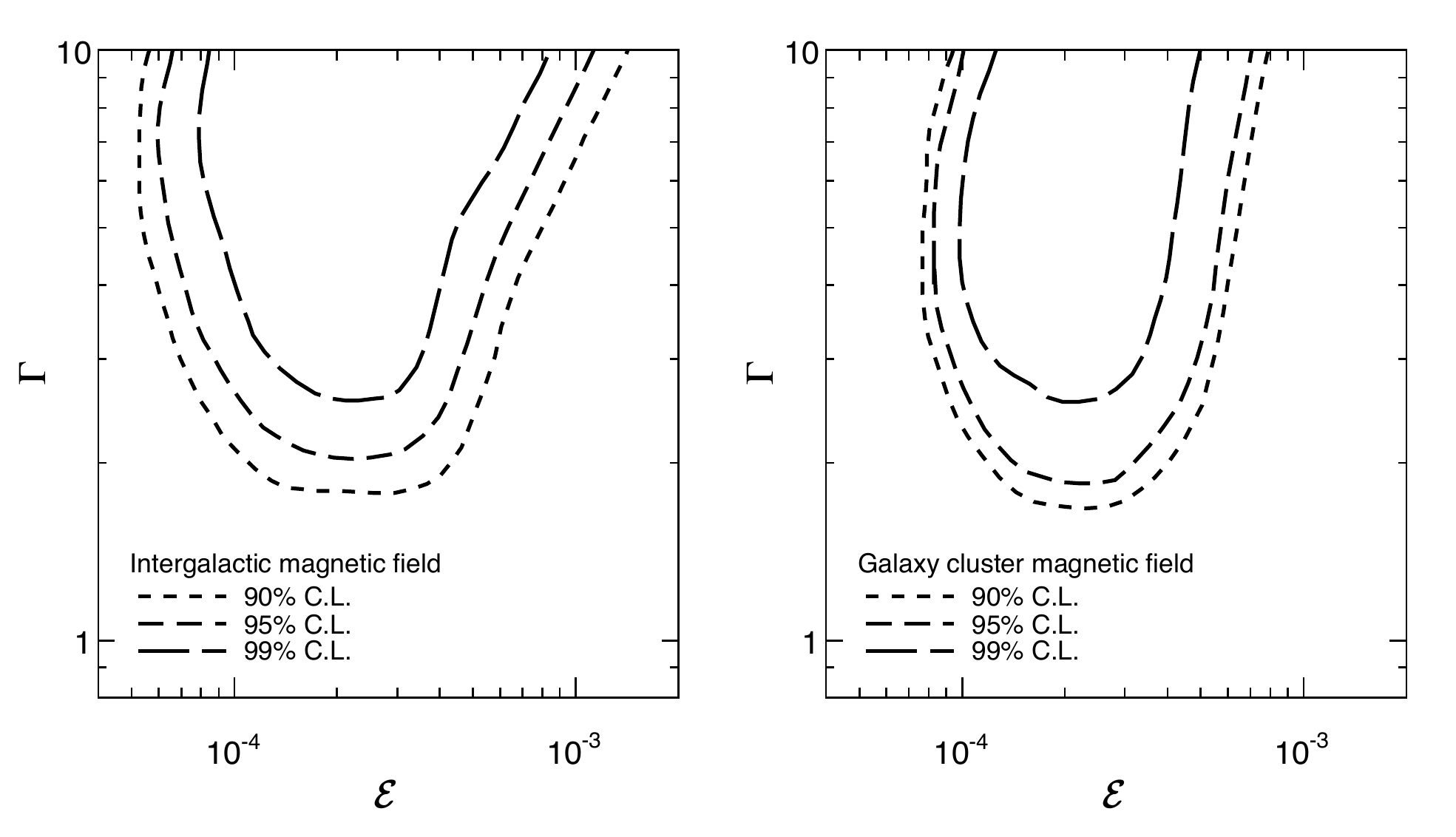}
\caption{Constraints on the ALP parameters expressed in reduced variables independent of the magnetic field strength (see text for details) for both the IGMF (left panel) and the galaxy cluster magnetic field (right panel).
\label{fig:constr1}}
\end{figure*}

It is interesting to derive constraints that do not explicitly depend on the magnetic field. This step is not necessary to derive the final limits, but it allows one to constrain the ALP parameters for other magnetic field values, or update the constraints if measurements of the magnetic field in the PKS~2155$-$304 cluster become available. To do so, the constraints are expressed using the following dimensionless parameters $\Gamma$ and $\mathcal{E}$:
\begin{eqnarray}
\Gamma\;=\;\frac{gBL}{2\sqrt{L/s}} \;\;\; \text{and}\;\;\;
\mathcal{E} & = & \frac{m}{\sqrt{B}} \label{eq:conv_Eps}\;\;.
\end{eqnarray}
The corresponding constraints obtained from conversions in the galaxy cluster magnetic field or in the IGMF  are shown in Fig.~\ref{fig:constr1}, for different values of the confidence level. By construction, the constraints on $\Gamma$ are at the same level for both types of magnetic fields.  The differences in the shape of the constraints are due to the fact that the EBL absorption acts in addition to the photon-ALP oscillations in the case of the IGMF, and to the fact that the number of equivalent domains are different.

The constraints in the ALP parameter space $(g_{\gamma a}, m)$ are deduced from the constraints on  $\Gamma$ and $\mathcal{E}$ presuming some values for $L$, $L/s$ and $B$. The limits are derived conservatively assuming magnetic field strengths of 1 $\mu$G within the domains of the cluster magnetic field. Higher values would lead to better constraints. Concerning the conversion in the IGMF, the constraints are subject to much larger uncertainties, since magnetic field strengths and turbulence scales are poorly known.
For the IGMF the total length is the distance to the source and a coherence length at redshift $z$ of $1 \; {\rm Mpc}/(1+z)$ is assumed. These parameters are used to fill in Eq.~(\ref{eq:conv_Eps}) and are summarized in Table~\ref{tab:param}. Note that the uncertainty on the angular size of the galaxy cluster translates into a 5\% systematic uncertainty on the constraint.

\begin{table}[t]
\centering
\begin{tabular*}{\columnwidth}{ccc}
\hline
\hline 
	& \hspace{1cm} Cluster magnetic field 	& \hspace{1cm} IGMF		\\
$B$ 	& \hspace{1cm}  1 $\mu$G			& \hspace{1cm}  1 nG 		\\
$L$ 	& \hspace{1cm}  370 kpc				& \hspace{1cm}   500 Mpc 	\\
$L/s$ 	& \hspace{1cm}  37					& \hspace{1cm}   528		\\
\hline
\hline
\end{tabular*}
\caption{Parameters used to fill in Eq.~(\ref{eq:conv_Eps}) to express the final constraints on the ALP parameters.\label{tab:param}}
\end{table}

\begin{figure*}
\centering
\includegraphics[width=\textwidth]{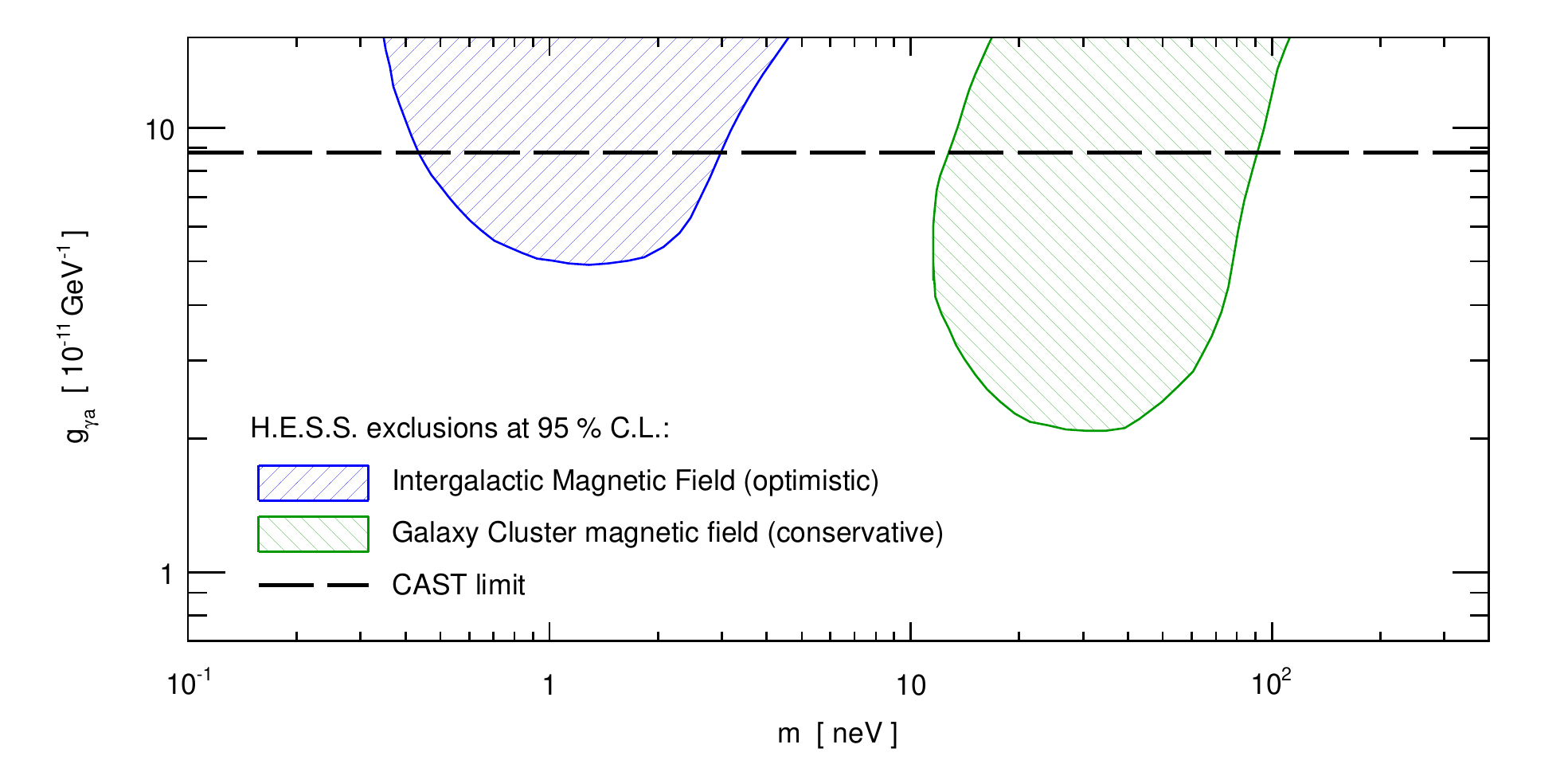}
\caption{H.E.S.S. exclusion limits on the ALP parameters $g_{\gamma a}$ and $m$. The dashed region on the left is obtained considering photon-ALP mixing in the IGMF with an optimistic scenario with a 1 nG field strength. The dashed region on the right is obtained considering photon-ALP mixing in the galaxy cluster of PKS~2155$-$304.
\label{fig:constr2}}
\end{figure*}

The obtained limits are displayed in Fig.~\ref{fig:constr2} for the conversion in the IGMF and for the conversion in the cluster magnetic field. 
 As anticipated in Sec.~\ref{sec:signal}, the H.E.S.S. limits peak at 1 neV in the case of IGMF conversions, and at 20 neV in the case of conversions in the cluster. In the case of the IGMF, the uncertainty on the strength of the magnetic field implies a range of possible constraints on $g_{\gamma a}$  between $10^{-11}$ and $10^{-3}\; \rm GeV^{-1}$. The limit that appears in Fig.~\ref{fig:constr2} is expressed for an IGMF strength of 1 nG. It is therefore optimistic. On the other hand, because of the observation of the galaxy cluster around PKS~2155$-$304, the conservatively value of 1 $\mu$G for its magnetic field and the estimator with minimal assumptions used here, the constraints obtained from the galaxy cluster are considered as robust.

The limits are of the order of a few $10^{-11}\;\rm GeV^{-1}$, improving the current CAST limit, which is $8.8\times 10^{-11}\;\rm GeV^{-1}$ in the mass range from $9\times10^{-9}$ to $10^{-7}\;\rm eV$. In the same region of the parameter space, other constraints come from the absence of a gamma-ray emission in coincidence with the SN 1987A neutrino burst. They apply for  ALP masses lower than 1 neV and restrict the coupling to values lower than $10^{-11}\;\rm GeV^{-1}$~\cite{Brockway:1996yr, Grifols:1996id}. In~\cite{2013PhRvD..87c5027M}, the authors computed the regions that would be allowed if ALPs were at the origin of an excess of transparency of the Universe. It should be noted that the present approach is complementary as it restricts the allowed ALP parameter space by an independent method. At even lower masses below $10^{-11}\;\rm eV$, an irregularity measurement based method applied to x-ray data yields a limit on the coupling of $8\times 10^{-12}\;\rm GeV^{-1}$~\cite{2013arXiv1304.0989W}. In the future, laboratory experiments such as IAXO~\cite{Vogel:2013bta} and ALPS~II~\cite{2013arXiv1302.5647B} should be sensitive to low-mass ALPs with couplings as low as $3\times10^{-12}\;\rm GeV^{-1}$.

The limits derived here for pseudoscalar particles are also valid for scalar particles. Indeed in the latter case, the term $\vec{E}\cdot\vec{B}$ entering Eq.~\ref{eq:lagrangian2} has to be replaced by $B^2$, implying that different polarization components of the photon are involved in the mixing. However, this would not influence the present analysis as the polarization components involved in the mixing are randomized together with the realizations of the magnetic field, and in particular its orientation. Therefore the limits presented in Figs.~\ref{fig:constr1}~and~\ref{fig:constr2} are directly applicable to the case of light scalar bosons that couple to photons.

\section{Summary}
\label{sec:summary}

Photon-ALP mixing in astrophysical sources is expected to manifest itself through the induction of irregularities in the energy spectra of high-energy gamma-ray sources. In this paper, H.E.S.S. observations of the BL Lac object PKS~2155$-$304 are used to derive constraints on the coupling strength of ALPs. In an optimistic scenario for the intergalactic magnetic field,  an upper limit of $5\times 10^{-11}\;\rm GeV^{-1}$ for the ALP coupling to photons is derived for ALPs of masses of order 1 neV. A conservative limit of $2.1\times 10^{-11}\;\rm GeV^{-1}$ is found for ALP masses around 25 neV when considering the galaxy cluster magnetic field. These results depend on assumptions on the magnetic field around the source, which are chosen conservatively.

These are the first exclusions on ALP mass and coupling to photons from gamma-ray astronomy, they improve the CAST constraints in this mass range by a factor of about 4. These limits are also valid for scalar particles.
In the future this method can be applied to observations including the fifth telescope of H.E.S.S., thus lowering energy threshold and widening the accessible ALP mass range. Other sources with different magnetic field turbulence configurations may be used as well. Altogether this will improve the sensitivity of this type of analysis, that could lead to an improvement of the limits or possibly a discovery.

\begin{acknowledgements}
  The support of the Namibian authorities and of the University of Namibia in facilitating the construction and operation of H.E.S.S. is gratefully acknowledged, as is the support by the German Ministry for Education and Research (BMBF), the Max Planck Society, the German Research Foundation (DFG), the French Ministry for Research, the CNRS-IN2P3 and the Astroparticle Interdisciplinary Program of the CNRS, the United Kingdom Science and Technology Facilities Council (STFC), the IPNP of the Charles University, the Czech Science Foundation, the Polish Ministry of Science and  Higher Education, the South African Department of Science and Technology and National Research Foundation, and the University of Namibia. We appreciate the excellent work of the technical support staff in Berlin, Durham, Hamburg, Heidelberg, Palaiseau, Paris, Saclay, and Namibia in the construction and operation of the equipment.
\end{acknowledgements}

\bibliography{axions_v7.0}

\begin{thebibliography}{72}
\expandafter\ifx\csname natexlab\endcsname\relax\def\natexlab#1{#1}\fi
\expandafter\ifx\csname bibnamefont\endcsname\relax
  \def\bibnamefont#1{#1}\fi
\expandafter\ifx\csname bibfnamefont\endcsname\relax
  \def\bibfnamefont#1{#1}\fi
\expandafter\ifx\csname citenamefont\endcsname\relax
  \def\citenamefont#1{#1}\fi
\expandafter\ifx\csname url\endcsname\relax
  \def\url#1{\texttt{#1}}\fi
\expandafter\ifx\csname urlprefix\endcsname\relax\def\urlprefix{URL }\fi
\providecommand{\bibinfo}[2]{#2}
\providecommand{\eprint}[2][]{\url{#2}}

\bibitem[{\citenamefont{Peccei and Quinn}(1977)}]{Peccei:1977hh}
\bibinfo{author}{\bibfnamefont{R.~D.} \bibnamefont{Peccei}} \bibnamefont{and}
  \bibinfo{author}{\bibfnamefont{H.~R.} \bibnamefont{Quinn}},
  \bibinfo{journal}{Phys.Rev.Lett.} \textbf{\bibinfo{volume}{38}},
  \bibinfo{pages}{1440} (\bibinfo{year}{1977}).

\bibitem[{\citenamefont{Wilczek}(1978)}]{Wilczek:1977pj}
\bibinfo{author}{\bibfnamefont{F.}~\bibnamefont{Wilczek}},
  \bibinfo{journal}{Phys.Rev.Lett.} \textbf{\bibinfo{volume}{40}},
  \bibinfo{pages}{279} (\bibinfo{year}{1978}).

\bibitem[{\citenamefont{Weinberg}(1978)}]{Weinberg:1977ma}
\bibinfo{author}{\bibfnamefont{S.}~\bibnamefont{Weinberg}},
  \bibinfo{journal}{Phys.Rev.Lett.} \textbf{\bibinfo{volume}{40}},
  \bibinfo{pages}{223} (\bibinfo{year}{1978}).

\bibitem[{\citenamefont{Kim}(1987)}]{Kim:1986ax}
\bibinfo{author}{\bibfnamefont{J.~E.} \bibnamefont{Kim}},
  \bibinfo{journal}{Phys.Rep.} \textbf{\bibinfo{volume}{150}},
  \bibinfo{pages}{1} (\bibinfo{year}{1987}).

\bibitem[{\citenamefont{{Raffelt}}(1990)}]{1990PhR...198....1R}
\bibinfo{author}{\bibfnamefont{G.~G.} \bibnamefont{{Raffelt}}},
  \bibinfo{journal}{Phys.Rep.} \textbf{\bibinfo{volume}{198}},
  \bibinfo{pages}{1} (\bibinfo{year}{1990}).

\bibitem[{\citenamefont{Svrcek and Witten}(2006)}]{Svrcek:2006yi}
\bibinfo{author}{\bibfnamefont{P.}~\bibnamefont{Svrcek}} \bibnamefont{and}
  \bibinfo{author}{\bibfnamefont{E.}~\bibnamefont{Witten}},
  \bibinfo{journal}{J. High Energy Phys.} \textbf{\bibinfo{volume}{06}},
  \bibinfo{pages}{051} (\bibinfo{year}{2006}), \eprint{arXiv: hep-th/0605206}.

\bibitem[{\citenamefont{Arvanitaki et~al.}(2010)\citenamefont{Arvanitaki,
  Dimopoulos, Dubovsky, Kaloper, and March-Russell}}]{Arvanitaki:2009fg}
\bibinfo{author}{\bibfnamefont{A.}~\bibnamefont{Arvanitaki}},
  \bibinfo{author}{\bibfnamefont{S.}~\bibnamefont{Dimopoulos}},
  \bibinfo{author}{\bibfnamefont{S.}~\bibnamefont{Dubovsky}},
  \bibinfo{author}{\bibfnamefont{N.}~\bibnamefont{Kaloper}}, \bibnamefont{and}
  \bibinfo{author}{\bibfnamefont{J.}~\bibnamefont{March-Russell}},
  \bibinfo{journal}{Phys.Rev.D} \textbf{\bibinfo{volume}{81}},
  \bibinfo{pages}{123530} (\bibinfo{year}{2010}), \eprint{arXiv:
  hep-th/0905.4720}.

\bibitem[{\citenamefont{Arias et~al.}(2012)}]{Arias:2012az}
\bibinfo{author}{\bibfnamefont{P.}~\bibnamefont{Arias}} \bibnamefont{et~al.},
  \bibinfo{journal}{J. Cosmol. Astropart. Phys.} \textbf{\bibinfo{volume}{06}},
  \bibinfo{pages}{013} (\bibinfo{year}{2012}), \eprint{arXiv:
  hep-ph/1201.5902}.

\bibitem[{\citenamefont{{Raffelt} and {Stodolsky}}(1988)}]{1988PhRvD..37.1237R}
\bibinfo{author}{\bibfnamefont{G.}~\bibnamefont{{Raffelt}}} \bibnamefont{and}
  \bibinfo{author}{\bibfnamefont{L.}~\bibnamefont{{Stodolsky}}},
  \bibinfo{journal}{Phys.Rev.D} \textbf{\bibinfo{volume}{37}},
  \bibinfo{pages}{1237} (\bibinfo{year}{1988}).

\bibitem[{\citenamefont{{Sikivie}}(1983)}]{1983PhRvL..51.1415S}
\bibinfo{author}{\bibfnamefont{P.}~\bibnamefont{{Sikivie}}},
  \bibinfo{journal}{Phys.Rev.Lett.} \textbf{\bibinfo{volume}{51}},
  \bibinfo{pages}{1415} (\bibinfo{year}{1983}).

\bibitem[{\citenamefont{Beringer et~al.}(2012)}]{Beringer:1900zz}
\bibinfo{author}{\bibfnamefont{J.}~\bibnamefont{Beringer}} \bibnamefont{et~al.}
  (\bibinfo{collaboration}{Particle Data Group}), \bibinfo{journal}{Phys.Rev.D}
  \textbf{\bibinfo{volume}{86}}, \bibinfo{pages}{010001}
  (\bibinfo{year}{2012}).

\bibitem[{\citenamefont{Arik et~al.}(2011)}]{2011PhRvL.107z1302A}
\bibinfo{author}{\bibfnamefont{M.}~\bibnamefont{Arik}} \bibnamefont{et~al.}
  (\bibinfo{collaboration}{CAST Collaboration}),
  \bibinfo{journal}{Phys.Rev.Lett.} \textbf{\bibinfo{volume}{107}},
  \bibinfo{eid}{261302} (\bibinfo{year}{2011}).

\bibitem[{\citenamefont{Asztalos et~al.}(2010)}]{Asztalos:2009yp}
\bibinfo{author}{\bibfnamefont{S.}~\bibnamefont{Asztalos}} \bibnamefont{et~al.}
  (\bibinfo{collaboration}{ADMX Collaboration}),
  \bibinfo{journal}{Phys.Rev.Lett.} \textbf{\bibinfo{volume}{104}},
  \bibinfo{pages}{041301} (\bibinfo{year}{2010}), \eprint{arXiv:
  astro-ph/0910.5914}.

\bibitem[{\citenamefont{Ehret et~al.}(2010)}]{Ehret:2010mh}
\bibinfo{author}{\bibfnamefont{K.}~\bibnamefont{Ehret}} \bibnamefont{et~al.}
  (\bibinfo{collaboration}{ALPS Collaboration}), \bibinfo{journal}{Phys.Lett.B}
  \textbf{\bibinfo{volume}{689}}, \bibinfo{pages}{149} (\bibinfo{year}{2010}),
  \eprint{arXiv: hep-ex/1004.1313}.

\bibitem[{\citenamefont{{Jaeckel} and {Ringwald}}(2010)}]{2010ARNPS..60..405J}
\bibinfo{author}{\bibfnamefont{J.}~\bibnamefont{{Jaeckel}}} \bibnamefont{and}
  \bibinfo{author}{\bibfnamefont{A.}~\bibnamefont{{Ringwald}}},
  \bibinfo{journal}{Annu.Rev.Nucl.Part.Sci} \textbf{\bibinfo{volume}{60}},
  \bibinfo{pages}{405} (\bibinfo{year}{2010}), \eprint{1002.0329}.

\bibitem[{\citenamefont{{Hooper} and {Serpico}}(2007)}]{2007PhRvL..99w1102H}
\bibinfo{author}{\bibfnamefont{D.}~\bibnamefont{{Hooper}}} \bibnamefont{and}
  \bibinfo{author}{\bibfnamefont{P.~D.} \bibnamefont{{Serpico}}},
  \bibinfo{journal}{Phys.Rev.Lett.} \textbf{\bibinfo{volume}{99}},
  \bibinfo{eid}{231102} (\bibinfo{year}{2007}).

\bibitem[{\citenamefont{{Mirizzi} et~al.}(2007)\citenamefont{{Mirizzi},
  {Raffelt}, and {Serpico}}}]{2007PhRvD..76b3001M}
\bibinfo{author}{\bibfnamefont{A.}~\bibnamefont{{Mirizzi}}},
  \bibinfo{author}{\bibfnamefont{G.~G.} \bibnamefont{{Raffelt}}},
  \bibnamefont{and} \bibinfo{author}{\bibfnamefont{P.~D.}
  \bibnamefont{{Serpico}}}, \bibinfo{journal}{Phys.Rev.D}
  \textbf{\bibinfo{volume}{76}}, \bibinfo{eid}{023001} (\bibinfo{year}{2007}).

\bibitem[{\citenamefont{{Hochmuth} and {Sigl}}(2007)}]{2007PhRvD..76l3011H}
\bibinfo{author}{\bibfnamefont{K.~A.} \bibnamefont{{Hochmuth}}}
  \bibnamefont{and} \bibinfo{author}{\bibfnamefont{G.}~\bibnamefont{{Sigl}}},
  \bibinfo{journal}{Phys.Rev.D} \textbf{\bibinfo{volume}{76}},
  \bibinfo{eid}{123011} (\bibinfo{year}{2007}).

\bibitem[{\citenamefont{{De Angelis} et~al.}(2008)\citenamefont{{De Angelis},
  {Mansutti}, and {Roncadelli}}}]{2008PhLB..659..847D}
\bibinfo{author}{\bibfnamefont{A.}~\bibnamefont{{De Angelis}}},
  \bibinfo{author}{\bibfnamefont{O.}~\bibnamefont{{Mansutti}}},
  \bibnamefont{and}
  \bibinfo{author}{\bibfnamefont{M.}~\bibnamefont{{Roncadelli}}},
  \bibinfo{journal}{Phys.Lett.B} \textbf{\bibinfo{volume}{659}},
  \bibinfo{pages}{847} (\bibinfo{year}{2008}), \eprint{arXiv:
  astro-ph/0707.2695}.

\bibitem[{\citenamefont{{Gould} and
  {Schr{\'e}der}}(1967)}]{1967PhRv..155.1408G}
\bibinfo{author}{\bibfnamefont{R.~J.} \bibnamefont{{Gould}}} \bibnamefont{and}
  \bibinfo{author}{\bibfnamefont{G.~P.} \bibnamefont{{Schr{\'e}der}}},
  \bibinfo{journal}{Phys.Rev.} \textbf{\bibinfo{volume}{155}},
  \bibinfo{pages}{1408} (\bibinfo{year}{1967}).

\bibitem[{\citenamefont{{Dwek} and {Krennrich}}(2013)}]{2013APh....43..112D}
\bibinfo{author}{\bibfnamefont{E.}~\bibnamefont{{Dwek}}} \bibnamefont{and}
  \bibinfo{author}{\bibfnamefont{F.}~\bibnamefont{{Krennrich}}},
  \bibinfo{journal}{Astropart.Phys.} \textbf{\bibinfo{volume}{43}},
  \bibinfo{pages}{112} (\bibinfo{year}{2013}), \eprint{arXiv:
  astro-ph/1209.4661}.

\bibitem[{\citenamefont{{Cs{\'a}ki} et~al.}(2003)\citenamefont{{Cs{\'a}ki},
  {Kaloper}, {Peloso}, and {Terning}}}]{2003JCAP...05..005C}
\bibinfo{author}{\bibfnamefont{C.}~\bibnamefont{{Cs{\'a}ki}}},
  \bibinfo{author}{\bibfnamefont{N.}~\bibnamefont{{Kaloper}}},
  \bibinfo{author}{\bibfnamefont{M.}~\bibnamefont{{Peloso}}}, \bibnamefont{and}
  \bibinfo{author}{\bibfnamefont{J.}~\bibnamefont{{Terning}}},
  \bibinfo{journal}{J. Cosmol. Astropart. Phys.} \textbf{\bibinfo{volume}{05}},
  \bibinfo{pages}{005} (\bibinfo{year}{2003}).

\bibitem[{\citenamefont{{De Angelis} et~al.}(2007)\citenamefont{{De Angelis},
  {Roncadelli}, and {Mansutti}}}]{2007PhRvD..76l1301D}
\bibinfo{author}{\bibfnamefont{A.}~\bibnamefont{{De Angelis}}},
  \bibinfo{author}{\bibfnamefont{M.}~\bibnamefont{{Roncadelli}}},
  \bibnamefont{and}
  \bibinfo{author}{\bibfnamefont{O.}~\bibnamefont{{Mansutti}}},
  \bibinfo{journal}{Phys.Rev.D} \textbf{\bibinfo{volume}{76}},
  \bibinfo{eid}{121301} (\bibinfo{year}{2007}).

\bibitem[{\citenamefont{{Simet} et~al.}(2008)\citenamefont{{Simet}, {Hooper},
  and {Serpico}}}]{2008PhRvD..77f3001S}
\bibinfo{author}{\bibfnamefont{M.}~\bibnamefont{{Simet}}},
  \bibinfo{author}{\bibfnamefont{D.}~\bibnamefont{{Hooper}}}, \bibnamefont{and}
  \bibinfo{author}{\bibfnamefont{P.~D.} \bibnamefont{{Serpico}}},
  \bibinfo{journal}{Phys.Rev.D} \textbf{\bibinfo{volume}{77}},
  \bibinfo{eid}{063001} (\bibinfo{year}{2008}), \eprint{arXiv:
  astro-ph/0712.2825}.

\bibitem[{\citenamefont{Mirizzi and Montanino}(2009)}]{Mirizzi:2009aj}
\bibinfo{author}{\bibfnamefont{A.}~\bibnamefont{Mirizzi}} \bibnamefont{and}
  \bibinfo{author}{\bibfnamefont{D.}~\bibnamefont{Montanino}},
  \bibinfo{journal}{J. Cosmol. Astropart. Phys.} \textbf{\bibinfo{volume}{12}},
  \bibinfo{pages}{004} (\bibinfo{year}{2009}), \eprint{arXiv:
  astro-ph/0911.0015}.

\bibitem[{\citenamefont{{S{\'a}nchez-Conde}
  et~al.}(2009)\citenamefont{{S{\'a}nchez-Conde}, {Paneque}, {Bloom}, {Prada},
  and {Dom{\'{\i}}nguez}}}]{2009PhRvD..79l3511S}
\bibinfo{author}{\bibfnamefont{M.~A.} \bibnamefont{{S{\'a}nchez-Conde}}},
  \bibinfo{author}{\bibfnamefont{D.}~\bibnamefont{{Paneque}}},
  \bibinfo{author}{\bibfnamefont{E.}~\bibnamefont{{Bloom}}},
  \bibinfo{author}{\bibfnamefont{F.}~\bibnamefont{{Prada}}}, \bibnamefont{and}
  \bibinfo{author}{\bibfnamefont{A.}~\bibnamefont{{Dom{\'{\i}}nguez}}},
  \bibinfo{journal}{Phys.Rev.D} \textbf{\bibinfo{volume}{79}},
  \bibinfo{eid}{123511} (\bibinfo{year}{2009}).

\bibitem[{\citenamefont{{De Angelis} et~al.}(2009)\citenamefont{{De Angelis},
  {Mansutti}, {Persic}, and {Roncadelli}}}]{2009MNRAS.394L..21D}
\bibinfo{author}{\bibfnamefont{A.}~\bibnamefont{{De Angelis}}},
  \bibinfo{author}{\bibfnamefont{O.}~\bibnamefont{{Mansutti}}},
  \bibinfo{author}{\bibfnamefont{M.}~\bibnamefont{{Persic}}}, \bibnamefont{and}
  \bibinfo{author}{\bibfnamefont{M.}~\bibnamefont{{Roncadelli}}},
  \bibinfo{journal}{Mon. Not. R. Astron. Soc.} \textbf{\bibinfo{volume}{394}},
  \bibinfo{pages}{L21} (\bibinfo{year}{2009}), \eprint{arXiv:
  astro-ph/0807.4246}.

\bibitem[{\citenamefont{{De Angelis} et~al.}(2011)\citenamefont{{De Angelis},
  {Galanti}, and {Roncadelli}}}]{2011PhRvD..84j5030D}
\bibinfo{author}{\bibfnamefont{A.}~\bibnamefont{{De Angelis}}},
  \bibinfo{author}{\bibfnamefont{G.}~\bibnamefont{{Galanti}}},
  \bibnamefont{and}
  \bibinfo{author}{\bibfnamefont{M.}~\bibnamefont{{Roncadelli}}},
  \bibinfo{journal}{Phys.Rev.D} \textbf{\bibinfo{volume}{84}},
  \bibinfo{eid}{105030} (\bibinfo{year}{2011}), \eprint{arXiv:
  astro-ph/1106.1132}.

\bibitem[{\citenamefont{{Meyer} et~al.}(2013)\citenamefont{{Meyer}, {Horns},
  and {Raue}}}]{2013PhRvD..87c5027M}
\bibinfo{author}{\bibfnamefont{M.}~\bibnamefont{{Meyer}}},
  \bibinfo{author}{\bibfnamefont{D.}~\bibnamefont{{Horns}}}, \bibnamefont{and}
  \bibinfo{author}{\bibfnamefont{M.}~\bibnamefont{{Raue}}},
  \bibinfo{journal}{Phys.Rev.D} \textbf{\bibinfo{volume}{87}},
  \bibinfo{eid}{035027} (\bibinfo{year}{2013}), \eprint{arXiv:
  astro-ph/1302.1208}.

\bibitem[{\citenamefont{Wouters and Brun}(2012)}]{Wouters:2012qd}
\bibinfo{author}{\bibfnamefont{D.}~\bibnamefont{Wouters}} \bibnamefont{and}
  \bibinfo{author}{\bibfnamefont{P.}~\bibnamefont{Brun}},
  \bibinfo{journal}{Phys.Rev.D} \textbf{\bibinfo{volume}{86}},
  \bibinfo{pages}{043005} (\bibinfo{year}{2012}), \eprint{arXiv:
  astro-ph/1205.6428}.

\bibitem[{\citenamefont{Mirizzi et~al.}(2005)\citenamefont{Mirizzi, Raffelt,
  and Serpico}}]{Mirizzi:2005ng}
\bibinfo{author}{\bibfnamefont{A.}~\bibnamefont{Mirizzi}},
  \bibinfo{author}{\bibfnamefont{G.~G.} \bibnamefont{Raffelt}},
  \bibnamefont{and} \bibinfo{author}{\bibfnamefont{P.~D.}
  \bibnamefont{Serpico}}, \bibinfo{journal}{Phys.Rev.D}
  \textbf{\bibinfo{volume}{72}}, \bibinfo{pages}{023501}
  (\bibinfo{year}{2005}), \eprint{astro-ph/0506078}.

\bibitem[{\citenamefont{Galanti and Roncadelli}(2013)}]{Galanti:2013afa}
\bibinfo{author}{\bibfnamefont{G.}~\bibnamefont{Galanti}} \bibnamefont{and}
  \bibinfo{author}{\bibfnamefont{M.}~\bibnamefont{Roncadelli}}
  (\bibinfo{year}{2013}), \eprint{1305.2114}.

\bibitem[{\citenamefont{Aharonian
  et~al.}(2005{\natexlab{a}})}]{2005A&A...430..865A}
\bibinfo{author}{\bibfnamefont{F.}~\bibnamefont{Aharonian}}
  \bibnamefont{et~al.} (\bibinfo{collaboration}{HESS Collaboration}),
  \bibinfo{journal}{Astron. Astrophys.} \textbf{\bibinfo{volume}{430}},
  \bibinfo{pages}{865} (\bibinfo{year}{2005}{\natexlab{a}}), \eprint{arXiv:
  astro-ph/0411582}.

\bibitem[{\citenamefont{Aharonian
  et~al.}(2005{\natexlab{b}})}]{2005A&A...442..895A}
\bibinfo{author}{\bibfnamefont{F.}~\bibnamefont{Aharonian}}
  \bibnamefont{et~al.} (\bibinfo{collaboration}{HESS Collaboration}),
  \bibinfo{journal}{Astron. Astrophys.} \textbf{\bibinfo{volume}{442}},
  \bibinfo{pages}{895} (\bibinfo{year}{2005}{\natexlab{b}}), \eprint{arXiv:
  astro-ph/0506593}.

\bibitem[{\citenamefont{Aharonian
  et~al.}(2009{\natexlab{a}})}]{2009ApJ...696L.150A}
\bibinfo{author}{\bibfnamefont{F.}~\bibnamefont{Aharonian}}
  \bibnamefont{et~al.} (\bibinfo{collaboration}{HESS Collaboration}),
  \bibinfo{journal}{Astrophys. J. Lett.} \textbf{\bibinfo{volume}{696}},
  \bibinfo{pages}{L150} (\bibinfo{year}{2009}{\natexlab{a}}), \eprint{arXiv:
  astro-ph/0903.2924}.

\bibitem[{\citenamefont{Aharonian et~al.}(2007)}]{2007ApJ...664L..71A}
\bibinfo{author}{\bibfnamefont{F.}~\bibnamefont{Aharonian}}
  \bibnamefont{et~al.} (\bibinfo{collaboration}{HESS Collaboration}),
  \bibinfo{journal}{Astrophys. J. Lett.} \textbf{\bibinfo{volume}{664}},
  \bibinfo{pages}{L71} (\bibinfo{year}{2007}), \eprint{arXiv:
  astro-ph/0706.0797}.

\bibitem[{\citenamefont{Aharonian
  et~al.}(2009{\natexlab{b}})}]{2009A&A...502..749A}
\bibinfo{author}{\bibfnamefont{F.}~\bibnamefont{Aharonian}}
  \bibnamefont{et~al.} (\bibinfo{collaboration}{HESS Collaboration}),
  \bibinfo{journal}{Astron. Astrophys.} \textbf{\bibinfo{volume}{502}},
  \bibinfo{pages}{749} (\bibinfo{year}{2009}{\natexlab{b}}), \eprint{arXiv:
  astro-ph/0906.2002}.

\bibitem[{\citenamefont{Abramowski et~al.}(2010)}]{2010A&A...520A..83H}
\bibinfo{author}{\bibfnamefont{A.}~\bibnamefont{Abramowski}}
  \bibnamefont{et~al.} (\bibinfo{collaboration}{HESS Collaboration}),
  \bibinfo{journal}{Astron. Astrophys.} \textbf{\bibinfo{volume}{520}},
  \bibinfo{eid}{A83} (\bibinfo{year}{2010}), \eprint{arXiv:
  astro-ph/1005.3702}.

\bibitem[{\citenamefont{{Smith} et~al.}(1995)\citenamefont{{Smith}, {O'Dea},
  and {Baum}}}]{1995ApJ...441..113S}
\bibinfo{author}{\bibfnamefont{E.~P.} \bibnamefont{{Smith}}},
  \bibinfo{author}{\bibfnamefont{C.~P.} \bibnamefont{{O'Dea}}},
  \bibnamefont{and} \bibinfo{author}{\bibfnamefont{S.~A.}
  \bibnamefont{{Baum}}}, \bibinfo{journal}{\apj}
  \textbf{\bibinfo{volume}{441}}, \bibinfo{pages}{113} (\bibinfo{year}{1995}).

\bibitem[{\citenamefont{Falomo et~al.}(1993)\citenamefont{Falomo, Pesce, and
  Treves}}]{Falomo:1993dv}
\bibinfo{author}{\bibfnamefont{R.}~\bibnamefont{Falomo}},
  \bibinfo{author}{\bibfnamefont{J.~E.} \bibnamefont{Pesce}}, \bibnamefont{and}
  \bibinfo{author}{\bibfnamefont{A.}~\bibnamefont{Treves}},
  \bibinfo{journal}{Astrophys.J.} \textbf{\bibinfo{volume}{411}},
  \bibinfo{pages}{L63} (\bibinfo{year}{1993}).

\bibitem[{\citenamefont{Ade et~al.}(2013)}]{2013arXiv1303.5076P}
\bibinfo{author}{\bibfnamefont{P.~A.~R.} \bibnamefont{Ade}}
  \bibnamefont{et~al.} (\bibinfo{collaboration}{Planck Collaboration})
  (\bibinfo{year}{2013}), \eprint{arXiv: astro-ph/1303.5076}.

\bibitem[{\citenamefont{{Carilli} and {Taylor}}(2002)}]{2002ARA&A..40..319C}
\bibinfo{author}{\bibfnamefont{C.~L.} \bibnamefont{{Carilli}}}
  \bibnamefont{and} \bibinfo{author}{\bibfnamefont{G.~B.}
  \bibnamefont{{Taylor}}}, \bibinfo{journal}{Annu.Rev.Astron.Astrophys.}
  \textbf{\bibinfo{volume}{40}}, \bibinfo{pages}{319} (\bibinfo{year}{2002}),
  \eprint{arXiv: astro-ph/0110655}.

\bibitem[{\citenamefont{Horns et~al.}(2012)}]{Horns:2012kw}
\bibinfo{author}{\bibfnamefont{D.}~\bibnamefont{Horns}} \bibnamefont{et~al.},
  \bibinfo{journal}{Phys.Rev.D} \textbf{\bibinfo{volume}{86}},
  \bibinfo{pages}{075024} (\bibinfo{year}{2012}), \eprint{arXiv:
  astro-ph/1207.0776}.

\bibitem[{\citenamefont{Ryu et~al.}(2012)\citenamefont{Ryu, Schleicher,
  Treumann, Tsagas, and Widrow}}]{Ryu:2011hu}
\bibinfo{author}{\bibfnamefont{D.}~\bibnamefont{Ryu}},
  \bibinfo{author}{\bibfnamefont{D.~R.} \bibnamefont{Schleicher}},
  \bibinfo{author}{\bibfnamefont{R.~A.} \bibnamefont{Treumann}},
  \bibinfo{author}{\bibfnamefont{C.~G.} \bibnamefont{Tsagas}},
  \bibnamefont{and} \bibinfo{author}{\bibfnamefont{L.~M.}
  \bibnamefont{Widrow}}, \bibinfo{journal}{Space Sci.Rev.}
  \textbf{\bibinfo{volume}{166}}, \bibinfo{pages}{1} (\bibinfo{year}{2012}),
  \eprint{arXiv: astro-ph/1109.4055}.

\bibitem[{\citenamefont{Akahori and Ryu}(2010)}]{Akahori:2010ym}
\bibinfo{author}{\bibfnamefont{T.}~\bibnamefont{Akahori}} \bibnamefont{and}
  \bibinfo{author}{\bibfnamefont{D.}~\bibnamefont{Ryu}},
  \bibinfo{journal}{Astrophys.J.} \textbf{\bibinfo{volume}{723}},
  \bibinfo{pages}{476} (\bibinfo{year}{2010}), \eprint{arXiv:
  astro-ph/1009.0570}.

\bibitem[{\citenamefont{{Neronov} and {Vovk}}(2010)}]{2010Sci...328...73N}
\bibinfo{author}{\bibfnamefont{A.}~\bibnamefont{{Neronov}}} \bibnamefont{and}
  \bibinfo{author}{\bibfnamefont{I.}~\bibnamefont{{Vovk}}},
  \bibinfo{journal}{Science} \textbf{\bibinfo{volume}{328}},
  \bibinfo{pages}{73} (\bibinfo{year}{2010}), \eprint{arXiv:
  astro-ph/1006.3504}.

\bibitem[{\citenamefont{{Dermer} et~al.}(2011)\citenamefont{{Dermer},
  {Cavadini}, {Razzaque}, {Finke}, {Chiang}, and {Lott}}}]{2011ApJ...733L..21D}
\bibinfo{author}{\bibfnamefont{C.~D.} \bibnamefont{{Dermer}}},
  \bibinfo{author}{\bibfnamefont{M.}~\bibnamefont{{Cavadini}}},
  \bibinfo{author}{\bibfnamefont{S.}~\bibnamefont{{Razzaque}}},
  \bibinfo{author}{\bibfnamefont{J.~D.} \bibnamefont{{Finke}}},
  \bibinfo{author}{\bibfnamefont{J.}~\bibnamefont{{Chiang}}}, \bibnamefont{and}
  \bibinfo{author}{\bibfnamefont{B.}~\bibnamefont{{Lott}}},
  \bibinfo{journal}{Astrophys.J.Lett.} \textbf{\bibinfo{volume}{733}},
  \bibinfo{eid}{L21} (\bibinfo{year}{2011}), \eprint{1011.6660}.

\bibitem[{\citenamefont{{Taylor} et~al.}(2011)\citenamefont{{Taylor}, {Vovk},
  and {Neronov}}}]{2011A&A...529A.144T}
\bibinfo{author}{\bibfnamefont{A.~M.} \bibnamefont{{Taylor}}},
  \bibinfo{author}{\bibfnamefont{I.}~\bibnamefont{{Vovk}}}, \bibnamefont{and}
  \bibinfo{author}{\bibfnamefont{A.}~\bibnamefont{{Neronov}}},
  \bibinfo{journal}{Astron. \& Astrophys} \textbf{\bibinfo{volume}{529}},
  \bibinfo{eid}{A144} (\bibinfo{year}{2011}), \eprint{1101.0932}.

\bibitem[{\citenamefont{Arlen et~al.}(2012)\citenamefont{Arlen, Vassiliev,
  Weisgarber, Wakely, and Shafi}}]{Arlen:2012iy}
\bibinfo{author}{\bibfnamefont{T.~C.} \bibnamefont{Arlen}},
  \bibinfo{author}{\bibfnamefont{V.~V.} \bibnamefont{Vassiliev}},
  \bibinfo{author}{\bibfnamefont{T.}~\bibnamefont{Weisgarber}},
  \bibinfo{author}{\bibfnamefont{S.~P.} \bibnamefont{Wakely}},
  \bibnamefont{and} \bibinfo{author}{\bibfnamefont{S.~Y.} \bibnamefont{Shafi}}
  (\bibinfo{year}{2012}), \eprint{arXiv: astro-ph/1210.2802}.

\bibitem[{\citenamefont{{Broderick} et~al.}(2012)\citenamefont{{Broderick},
  {Chang}, and {Pfrommer}}}]{2012ApJ...752...22B}
\bibinfo{author}{\bibfnamefont{A.~E.} \bibnamefont{{Broderick}}},
  \bibinfo{author}{\bibfnamefont{P.}~\bibnamefont{{Chang}}}, \bibnamefont{and}
  \bibinfo{author}{\bibfnamefont{C.}~\bibnamefont{{Pfrommer}}},
  \bibinfo{journal}{Astrophys.J.} \textbf{\bibinfo{volume}{752}},
  \bibinfo{eid}{22} (\bibinfo{year}{2012}), \eprint{arXiv: astro-ph/1106.5494}.

\bibitem[{\citenamefont{{Schlickeiser}
  et~al.}(2012)\citenamefont{{Schlickeiser}, {Ibscher}, and
  {Supsar}}}]{2012ApJ...758..102S}
\bibinfo{author}{\bibfnamefont{R.}~\bibnamefont{{Schlickeiser}}},
  \bibinfo{author}{\bibfnamefont{D.}~\bibnamefont{{Ibscher}}},
  \bibnamefont{and} \bibinfo{author}{\bibfnamefont{M.}~\bibnamefont{{Supsar}}},
  \bibinfo{journal}{Astrophys.J.} \textbf{\bibinfo{volume}{758}},
  \bibinfo{eid}{102} (\bibinfo{year}{2012}).

\bibitem[{\citenamefont{{Blasi} et~al.}(1999)\citenamefont{{Blasi}, {Burles},
  and {Olinto}}}]{1999ApJ...514L..79B}
\bibinfo{author}{\bibfnamefont{P.}~\bibnamefont{{Blasi}}},
  \bibinfo{author}{\bibfnamefont{S.}~\bibnamefont{{Burles}}}, \bibnamefont{and}
  \bibinfo{author}{\bibfnamefont{A.~V.} \bibnamefont{{Olinto}}},
  \bibinfo{journal}{Astrophys.J.Lett.} \textbf{\bibinfo{volume}{514}},
  \bibinfo{pages}{L79} (\bibinfo{year}{1999}), \eprint{arXiv:
  astro-ph/9812487}.

\bibitem[{\citenamefont{{Durrer} and {Neronov}}(2013)}]{2013A&ARv..21...62D}
\bibinfo{author}{\bibfnamefont{R.}~\bibnamefont{{Durrer}}} \bibnamefont{and}
  \bibinfo{author}{\bibfnamefont{A.}~\bibnamefont{{Neronov}}},
  \bibinfo{journal}{Astron. Astrophys. Rev.} \textbf{\bibinfo{volume}{21}},
  \bibinfo{pages}{62} (\bibinfo{year}{2013}), \eprint{arXiv:
  astro-ph/1303.7121}.

\bibitem[{\citenamefont{{Giacalone} and {Jokipii}}(1999)}]{1999ApJ...520..204G}
\bibinfo{author}{\bibfnamefont{J.}~\bibnamefont{{Giacalone}}} \bibnamefont{and}
  \bibinfo{author}{\bibfnamefont{J.~R.} \bibnamefont{{Jokipii}}},
  \bibinfo{journal}{Astrophys.J.} \textbf{\bibinfo{volume}{520}},
  \bibinfo{pages}{204} (\bibinfo{year}{1999}).

\bibitem[{\citenamefont{{Bonafede} et~al.}(2010)\citenamefont{{Bonafede},
  {Feretti}, {Murgia}, {Govoni}, {Giovannini}, {Dallacasa}, {Dolag}, and
  {Taylor}}}]{2010A&A...513A..30B}
\bibinfo{author}{\bibfnamefont{A.}~\bibnamefont{{Bonafede}}},
  \bibinfo{author}{\bibfnamefont{L.}~\bibnamefont{{Feretti}}},
  \bibinfo{author}{\bibfnamefont{M.}~\bibnamefont{{Murgia}}},
  \bibinfo{author}{\bibfnamefont{F.}~\bibnamefont{{Govoni}}},
  \bibinfo{author}{\bibfnamefont{G.}~\bibnamefont{{Giovannini}}},
  \bibinfo{author}{\bibfnamefont{D.}~\bibnamefont{{Dallacasa}}},
  \bibinfo{author}{\bibfnamefont{K.}~\bibnamefont{{Dolag}}}, \bibnamefont{and}
  \bibinfo{author}{\bibfnamefont{G.~B.} \bibnamefont{{Taylor}}},
  \bibinfo{journal}{Astron. \& Astrophys.} \textbf{\bibinfo{volume}{513}},
  \bibinfo{eid}{A30} (\bibinfo{year}{2010}), \eprint{arXiv:
  astro-ph/1002.0594}.

\bibitem[{\citenamefont{{Vogt} and {En{\ss}lin}}(2005)}]{2005A&A...434...67V}
\bibinfo{author}{\bibfnamefont{C.}~\bibnamefont{{Vogt}}} \bibnamefont{and}
  \bibinfo{author}{\bibfnamefont{T.~A.} \bibnamefont{{En{\ss}lin}}},
  \bibinfo{journal}{Astron. \& Astrophys.} \textbf{\bibinfo{volume}{434}},
  \bibinfo{pages}{67} (\bibinfo{year}{2005}), \eprint{arXiv: astro-ph/0501211}.

\bibitem[{\citenamefont{{Laurent-Muehleisen}
  et~al.}(1993)\citenamefont{{Laurent-Muehleisen}, {Kollgaard}, {Moellenbrock},
  and {Feigelson}}}]{1993AJ....106..875L}
\bibinfo{author}{\bibfnamefont{S.~A.} \bibnamefont{{Laurent-Muehleisen}}},
  \bibinfo{author}{\bibfnamefont{R.~I.} \bibnamefont{{Kollgaard}}},
  \bibinfo{author}{\bibfnamefont{G.~A.} \bibnamefont{{Moellenbrock}}},
  \bibnamefont{and} \bibinfo{author}{\bibfnamefont{E.~D.}
  \bibnamefont{{Feigelson}}}, \bibinfo{journal}{Astron.J.}
  \textbf{\bibinfo{volume}{106}}, \bibinfo{pages}{875} (\bibinfo{year}{1993}).

\bibitem[{\citenamefont{{Grasso} and {Rubinstein}}(2001)}]{2001PhR...348..163G}
\bibinfo{author}{\bibfnamefont{D.}~\bibnamefont{{Grasso}}} \bibnamefont{and}
  \bibinfo{author}{\bibfnamefont{H.~R.} \bibnamefont{{Rubinstein}}},
  \bibinfo{journal}{Phys.Rep.} \textbf{\bibinfo{volume}{348}},
  \bibinfo{pages}{163} (\bibinfo{year}{2001}), \eprint{arXiv:
  astro-ph/0009061}.

\bibitem[{\citenamefont{Aharonian et~al.}(2006)}]{2006A&A...457..899A}
\bibinfo{author}{\bibfnamefont{F.}~\bibnamefont{Aharonian}}
  \bibnamefont{et~al.} (\bibinfo{collaboration}{HESS Collaboration}),
  \bibinfo{journal}{Astron. Astrophys.} \textbf{\bibinfo{volume}{457}},
  \bibinfo{pages}{899} (\bibinfo{year}{2006}), \eprint{arXiv:
  astro-ph/0607333}.

\bibitem[{\citenamefont{{de Naurois} and {Rolland}}(2009)}]{deNaurois:2009ud}
\bibinfo{author}{\bibfnamefont{M.}~\bibnamefont{{de Naurois}}}
  \bibnamefont{and}
  \bibinfo{author}{\bibfnamefont{L.}~\bibnamefont{{Rolland}}},
  \bibinfo{journal}{Astrop.Phys.} \textbf{\bibinfo{volume}{32}},
  \bibinfo{pages}{231} (\bibinfo{year}{2009}), \eprint{0907.2610}.

\bibitem[{\citenamefont{Albert et~al.}(2007)}]{Albert:2007qw}
\bibinfo{author}{\bibfnamefont{J.}~\bibnamefont{Albert}} \bibnamefont{et~al.}
  (\bibinfo{collaboration}{MAGIC Collaboration}), \bibinfo{journal}{Methods
  Phys. Res., Sect. A} \textbf{\bibinfo{volume}{A583}}, \bibinfo{pages}{494}
  (\bibinfo{year}{2007}), \eprint{arXiv: astro-ph/0707.2453}.

\bibitem[{\citenamefont{Piron et~al.}(2001)}]{2001A&A...374..895P}
\bibinfo{author}{\bibfnamefont{F.}~\bibnamefont{Piron}} \bibnamefont{et~al.},
  \bibinfo{journal}{Astron. \& Astrophys.} \textbf{\bibinfo{volume}{374}},
  \bibinfo{pages}{895} (\bibinfo{year}{2001}), \eprint{arXiv:
  astro-ph/0106196}.

\bibitem[{\citenamefont{{Franceschini}
  et~al.}(2008)\citenamefont{{Franceschini}, {Rodighiero}, and
  {Vaccari}}}]{2008A&A...487..837F}
\bibinfo{author}{\bibfnamefont{A.}~\bibnamefont{{Franceschini}}},
  \bibinfo{author}{\bibfnamefont{G.}~\bibnamefont{{Rodighiero}}},
  \bibnamefont{and}
  \bibinfo{author}{\bibfnamefont{M.}~\bibnamefont{{Vaccari}}},
  \bibinfo{journal}{Astron. \& Astrophys.} \textbf{\bibinfo{volume}{487}},
  \bibinfo{pages}{837} (\bibinfo{year}{2008}), \eprint{arXiv:
  astro-ph/0805.1841}.

\bibitem[{\citenamefont{Abramowski et~al.}(2013)}]{2013A&A...550A...4H}
\bibinfo{author}{\bibfnamefont{A.}~\bibnamefont{Abramowski}}
  \bibnamefont{et~al.} (\bibinfo{collaboration}{HESS Collaboration}),
  \bibinfo{journal}{Astron. Astrophys.} \textbf{\bibinfo{volume}{550}},
  \bibinfo{eid}{A4} (\bibinfo{year}{2013}), \eprint{arXiv: astro-ph/1212.3409}.

\bibitem[{\citenamefont{{Katarzy{\'n}ski}
  et~al.}(2001)\citenamefont{{Katarzy{\'n}ski}, {Sol}, and
  {Kus}}}]{2001A&A...367..809K}
\bibinfo{author}{\bibfnamefont{K.}~\bibnamefont{{Katarzy{\'n}ski}}},
  \bibinfo{author}{\bibfnamefont{H.}~\bibnamefont{{Sol}}}, \bibnamefont{and}
  \bibinfo{author}{\bibfnamefont{A.}~\bibnamefont{{Kus}}},
  \bibinfo{journal}{Astron. \& Astrophys.} \textbf{\bibinfo{volume}{367}},
  \bibinfo{pages}{809} (\bibinfo{year}{2001}).

\bibitem[{\citenamefont{Aharonian et~al.}(1999)}]{1999A&A...349...11A}
\bibinfo{author}{\bibfnamefont{F.}~\bibnamefont{Aharonian}}
  \bibnamefont{et~al.}, \bibinfo{journal}{Astron. Astrophys.}
  \textbf{\bibinfo{volume}{349}}, \bibinfo{pages}{11} (\bibinfo{year}{1999}),
  \eprint{arXiv: astro-ph/9903386}.

\bibitem[{\citenamefont{Aharonian et~al.}(2003)}]{2003A&A...403..523A}
\bibinfo{author}{\bibfnamefont{F.}~\bibnamefont{Aharonian}}
  \bibnamefont{et~al.}, \bibinfo{journal}{Astron. Astrophys.}
  \textbf{\bibinfo{volume}{403}}, \bibinfo{pages}{523} (\bibinfo{year}{2003}),
  \eprint{arXiv: astro-ph/0301437}.

\bibitem[{\citenamefont{Brockway et~al.}(1996)\citenamefont{Brockway, Carlson,
  and Raffelt}}]{Brockway:1996yr}
\bibinfo{author}{\bibfnamefont{J.~W.} \bibnamefont{Brockway}},
  \bibinfo{author}{\bibfnamefont{E.~D.} \bibnamefont{Carlson}},
  \bibnamefont{and} \bibinfo{author}{\bibfnamefont{G.~G.}
  \bibnamefont{Raffelt}}, \bibinfo{journal}{Phys.Lett.}
  \textbf{\bibinfo{volume}{B383}}, \bibinfo{pages}{439} (\bibinfo{year}{1996}),
  \eprint{arXiv: astro-ph/9605197}.

\bibitem[{\citenamefont{Grifols et~al.}(1996)\citenamefont{Grifols, Masso, and
  Toldra}}]{Grifols:1996id}
\bibinfo{author}{\bibfnamefont{J.}~\bibnamefont{Grifols}},
  \bibinfo{author}{\bibfnamefont{E.}~\bibnamefont{Masso}}, \bibnamefont{and}
  \bibinfo{author}{\bibfnamefont{R.}~\bibnamefont{Toldra}},
  \bibinfo{journal}{Phys.Rev.Lett.} \textbf{\bibinfo{volume}{77}},
  \bibinfo{pages}{2372} (\bibinfo{year}{1996}), \eprint{arXiv:
  astro-ph/9606028}.

\bibitem[{\citenamefont{{Wouters} and {Brun}}(2013)}]{2013arXiv1304.0989W}
\bibinfo{author}{\bibfnamefont{D.}~\bibnamefont{{Wouters}}} \bibnamefont{and}
  \bibinfo{author}{\bibfnamefont{P.}~\bibnamefont{{Brun}}},
  \bibinfo{journal}{Astrophys.J.} \textbf{\bibinfo{volume}{772}},
  \bibinfo{pages}{44} (\bibinfo{year}{2013}), \eprint{arXiv:
  astro-ph/1304.0989}.

\bibitem[{\citenamefont{Vogel et~al.}(2013)}]{Vogel:2013bta}
\bibinfo{author}{\bibfnamefont{J.}~\bibnamefont{Vogel}} \bibnamefont{et~al.}
  (\bibinfo{year}{2013}), \eprint{arXiv: physics.ins-det/1302.3273}.

\bibitem[{\citenamefont{B{\"a}hre et~al.}(2013)}]{2013arXiv1302.5647B}
\bibinfo{author}{\bibfnamefont{R.}~\bibnamefont{B{\"a}hre}}
  \bibnamefont{et~al.} (\bibinfo{collaboration}{ALPS Collaboration}),
  \bibinfo{journal}{ArXiv e-prints}  (\bibinfo{year}{2013}), \eprint{arXiv:
  physics.ins-det/1302.5647}.

\end{thebibliography}

\end{document}